\documentclass[pra,twocolumn,preprintnumbers,amsmath,amssymb,superscriptaddress,showpacs,longbibliography]{revtex4-1}
\usepackage{graphicx}
\usepackage{amsmath}
\usepackage{graphics}
\usepackage{amssymb}
\usepackage{amsfonts,epsfig}
\usepackage{natbib}
\usepackage{hyperref}
\usepackage{enumitem}
\usepackage{dsfont}

\usepackage{xcolor}

\newcommand{\ket}[1]{\left|#1\right>}
\newcommand{\bra}[1]{\left< #1 \right|}
\newcommand{\beq}{\begin{equation}}
\newcommand{\eeq}{\end{equation}}
\newcommand{\bea}{\begin{eqnarray}}
\newcommand{\eea}{\end{eqnarray}}

\newcommand{\HH}{\hat{H}}

\newcommand{\one}{\mathds{1}}
\newcommand{\Tr}{\mathrm{Tr}}

\usepackage{color}

\usepackage{ulem} 

\begin{document}

\title{Decoherence of two entangled spin qubits coupled to an interacting sparse nuclear spin bath:
application to nitrogen vacancy centers}
\author{Damian Kwiatkowski}
\email{kwiatkowski@ifpan.edu.pl}
\affiliation{Institute of Physics, Polish Academy of Sciences, al.~Lotnik{\'o}w 32/46, PL 02-668 Warszawa, Poland}
\author{{\L}ukasz Cywi{\'n}ski}
\email{lcyw@ifpan.edu.pl}
\affiliation{Institute of Physics, Polish Academy of Sciences, al.~Lotnik{\'o}w 32/46, PL 02-668 Warszawa, Poland}

\date{\today}

\begin{abstract}
We consider pure dephasing of Bell states of electron spin qubits interacting with a sparse bath of nuclear spins. Using the newly developed two-qubit generalization of cluster correlation expansion method, we calculate the spin echo decay of $|\Psi\rangle$ and $|\Phi\rangle$ states for various interqubit distances. Comparing the results with calculations in which dephasing of each qubit is treated independently, we identify signatures of influence of common part of the bath on the two qubits. At large interqubit distances, this common part consists of many nuclei weakly coupled to both qubits, so that decoherence caused by it can be modeled by considering multiple uncorrelated sources of noise (clusters of nuclei), each of them weakly affecting the qubits. Consequently, the resulting genuinely two-qubit contribution to decoherence can be described as being caused by classical Gaussian noise. On the other hand, for small interqubit distances the common part of the environment contains clusters of spins that are strongly coupled to both qubits, and their contribution to two-qubit dephasing has visibly non-Gaussian character. We show that one van easily obtain information about non-Gaussianity of environmental noise affecting the qubits from the comparison of dephasing of $|\Psi\rangle$ and $|\Phi\rangle$ Bell states.
Numerical results are obtained for two nitrogen vacancy centers interacting with a bath of $^{13}$C nuclei of natural concentration, for which we obtain that Gaussian description of correlated part of environmental noise starts to hold for centers separated by about 3 nm.
\end{abstract}


\maketitle

\section{Introduction}
Correlations in environmental noise experienced by an entangled pair of qubits strongly affect their decoherence \cite{Duan_PRA98,Yu_PRB03,Szankowski_QIP15,Szankowski_PRA16,Paz_PRA17}. The most well-known example is the fact that $\ket{\Psi_{\pm}}$ Bell states form a decoherence-free subspace \cite{Lidar_ACP14}  when both qubits are exposed to exactly the same phase noise.
 When the environment can be treated as a source of classical noise, one can investigate the full spectrum of cases, from perfectly correlated to completely independent noises, by assuming a given form of cross-correlation of the noises experienced by the two qubits \cite{Szankowski_PRA16}. However, when environment has to be treated fully quantum mechanically, only the limiting cases, such as the perfect corelation or perfect anticorrelation, are easily defined, and the treatment of the most realistic case of a {\it partially common} environment (with two qubits located at some distance one from another and interacting with the surrounding environment) requires using methods designed for specific kinds of qubit-environment systems. 

With multiple experimental platforms having achieved  or being close to achieving the stage of development at which  multi-qubit registers are experimentally investigated \cite{Barends_Nature14,Ballance_PRL16,Neill_Science18,Zajac_PRAPL16,Vandersypen_NPJQI17}, the problem of realistic description of mutli-qubit decoherence is a pressing one. If the qubits in a multi-qubit register are all subjected to completely independent local noises, then the available knowledge on single-qubit decoherence, acquired during the preceding research on these qubits, is enough to describe the multi-qubit decoherence and disentanglement. However, if decoherence of two qubits is in fact correlated, arising from interaction with a common part of environment coupled appreciably to both qubits, then new features arise. They must be understood in order to make progress on using multiple qubits for quantum computation or quantum metrology. For the former application, quantum error correction \cite{Devitt_RPP13,Brown_RMP16} methods will have to be used in order to perform useful algorithms - and it is well known that assumptions about presence or lack of correlations between errors experienced by qubits \cite{Ng_PRA09,Novais_PRL13,Hutter_PRA14,Preskill_QIC13} are crucially important for construction of error correction protocols. In the context of quantum metrology, long-distance noise correlations can help in achieving Heisenberg limit of sensing accuracy with entangled quantum probes \cite{Jeske_NJP14}.

In this work, we consider the influence of partially common environment on entanglement dynamics of two spin qubits. Specifically, we focus on a model directly relevant to nitrogen-vacancy (NV) centers in diamond \cite{Dobrovitski_ARCMP13,Rondin_RPP14}: that of an electron spin qubit coupled via dipolar interaction to a {\it sparse} bath of nuclear spins that are themselves coupled to each other by dipolar interaction. We consider the regime of magnetic fields for which the dipolar-induced dynamics of groups of nuclei is the cause of decoherence. 
This is an example of experimentally relevant model of open quantum system for which the decoherence is caused by nontrivial dynamics of the interacting environment that is amenable to a well-controlled approximate quantum mechanical treatment, the so-called cluster correlation expansion \cite{Yang_CCE_PRB08,Yang_CCE_PRB09,Zhao_PRB12},  shown to give results in very good agreement with experiments on single qubits \cite{Zhao_PRB12}.

Let us first establish the basic feature of electron spin qubit dephasing due to interaction with such a bath. For electron spin qubit interacting with a nuclear spin bath, the assumption of pure dephasing, i.e.~no exchange of energy between the qubit and the bath, is most often a very good one (with exception of spin-$1/2$ qubits at very low magnetic fields). The qubit-bath coupling is thus of the $\hat{S}_{z}\hat{V}$ form, with $\hat{S}_{z}$ being the $z$-component of electron's spin operator, and $\hat{V} \! \propto \sum_{k,j}\hat{I}^{k}_{j}$ with $k$ numbering the nuclei and $j\! =\! x$, $y$, $z$.  
In such a case, the decoherence occurs due to dynamics of $\hat{V}$ caused by the environment's self-Hamiltonian $\hat{H}_{E}$ that describes the Larmor precession of nuclei, and their mutual dipolar interactions -- and our focus here is on the effects of the latter.

The long-range dipolar nature of qubit-nuclei and internuclear interactions complicates the problem by making it impossible to establish characteristic length-scales. However, if we focus on description of decoherence on a certain finite timescale, e.g.~that of half-decay of qubits' spin echo signal, it turns out that one can take into account only a finite size of the bath, and an effectively finite range of internuclear interaction.
This is possible because of the weakness of inter-nuclear dipolar interaction compared to dipolar interaction of an {\it electron} spin qubit with a nuclear spin (for the same spin-spin distance the latter is $~\sim \! 10^3$ times larger). The intrabath dynamics is then expected to be slow compared to qubit's dynamics, so that  dynamical correlations between multiple nuclei do not have enough time to build up before the qubits dephase appreciably \footnote{Note that a similar cluster-based approach to a problem of decoherence due to bath of {\it electron} spins, for which there is no clear separation of timescales of qubit's and bath dynamics, is possible to a certain degree, but at the price of introduction of multiple cumbersome and numerically costly modifications, see \cite{Witzel_PRB12}.}. This expectation motivated development of approaches \cite{Yao_PRB06,Witzel_PRB06,Saikin_PRB07,Yang_CCE_PRB08,Yang_CCE_PRB09,Yang_RPP17}, in which electron spin decoherence is approximated by considering groups (``clusters'') of increasing size: single spins subjected to Larmor precession, pairs of near-neighbor spins coupled via dipolar interactions, clusters of three spins, etc. In the most transparent Cluster Correlation Expansion (CCE) method \cite{Yang_CCE_PRB08,Yang_CCE_PRB09,Zhao_PRB12,Yang_RPP17} decoherence is expressed as a product of irreducible contributions of these clusters, with maximal size of the cluster chosen to achieve convergence of coherence signal up to the chosen maximal time. The influence of the interacting nuclear bath is expressed in this way as being due to an action of many uncorrelated small quantum systems. When the influence that each of these systems exerts on the qubit is weak, and when there are many systems exerting similar influence, it is clear that the net effect that the environment has on the qubit is then equivalent to that of classical Gaussian noise.  
However, if there is a certain number of nuclear clusters strongly coupled to the qubits that need to be considered to correctly describe the time dependence of qubits' coherence, the classical Gaussian noise picture  breaks down \cite{Reinhard_PRL12,Hall_PRB14,Ma_PRB15}.

The last important feature of the bath considered here is its sparsity, by which we mean the situation in which the nuclear spins in a given bath have essentially random positions. Even if their locations are constrained to the position of atoms in a crystalline lattice, as it is the case for nuclear baths in semiconductors, when only a fraction of nuclei has nonzero spin (e.g.~only $1.1$\% of carbon nuclei are of the spinful $^{13}$C isotope in natural diamond), they can be considered to be randomly and uniformly distributed in space -- the underlying lattice is relevant only when considering nuclei very close (i.e.~few lattice sites away) to the qubit. An important feature of this kind of bath is that decoherence of a qubit can exhibit dependence on details of spatial configuration of nuclei -- in other words the decoherence signal in general depends on {\it spatial realization} of the bath.

In this paper we use the CCE method, adapt it to the calculation of {\it two-qubit coherence}, and apply it to the case of pure dephasing of Bell states of two qubits separated by a finite distance. We focus here on decoherence observed in spin echo experiment, in which both qubits are simultaneously subjected to a $\pi$ pulse at  midpoint of their evolution. In such an experiment we remove the influence of the slowest fluctuations of the bath (that can be described classically as nuclear spin diffusion \cite{Reilly_PRL08,Reilly_PRL10,Malinowski_PRL17}), and the observed decoherence is due to much faster dynamics of small groups of nuclei, for which the status of applicability of classical approximations is not settled in an obvious way. Specifically, we focus on regime of magnetic fields large enough to suppress single-nucleus contributions to decoherence, and we focus on two-qubit dephasing caused by dynamics of pairs (and larger clusters) of nuclei coupled by dipolar interactions.  
The two main results are (1) showing under which conditions the effects of common environment will be observable in diamond samples containing natural concentration of $^{13}$C nuclear spins, and (2) devising a simple way of detecting non-Gaussian statistic of the common noise affecting the qubits, that allows us to show when the influence of the common part of the environment can be treated using classical Gaussian approximation. 
These  results are relevant for understanding of prospects of efficient quantum error correction, as the available treatments of error correction protocols in presence of correlated errors rely on treating the environment as a source of Gaussian noise \cite{Ng_PRA09}. 
The latter one is also important for the research on using qubits as spectrometers of environmental noise, as the assumption of Gaussian nature of environmental noise underpins the widely used noise spectroscopy protocols based on dynamical decoupling of a single qubit (see \cite{Szankowski_JPCM17,Degen_RMP17} and references therein) and multiple qubits \cite{Szankowski_PRA16,Szankowski_JPCM17,Paz_PRA17}, while characterization of non-Gaussian features of environmental noise is a subject of ongoing theoretical investigations \cite{Ramon_PRB15,Norris_PRL16,Szankowski_JPCM17}.

The model that we consider here applies to the case of NV centers in diamond, and while we focus on this case, let us mention that the model is applicable to a wider class of electron spin qubits dipolarly coupled to sparse nuclear baths, that includes other spinful deep defects such as silicon vacancies and divacancies in SiC \cite{Widmann_NM15,Seo_NC16,Carter_PRB15}. We perform calculations for natural concentration of $^{13}$C spinful isotope in diamond lattice. Creation, control and readout of entangled states of two such NV center qubits qubits separated by $\approx 20$ nm distance was shown recently \cite{Dolde_NP13}, and the free-evolution (without echo pulses) dephasing of Bell states measured there showed no discernible signatures of common bath effects. 
Here we consider smaller interqubit distances ($d \! < \! 5$ nm), decoherence under two-qubit spin echo sequence, and parallel quantization axes, with magnetic field gradient allowing for separate addressing of the two qubits. The latter also suppresses the interqubit flip-flops due to their mutual dipolar interaction,  enabling the use of pure dephasing approximation, and allowing for straightforward generalization of CCE method to calculation of two-qubit coherence dynamics. We use magnetic field of $300$ mT, at which single-nucleus contributions to echo decay are strongly suppressed \cite{Zhao_PRB12}, and, as we checked, spin echo decoherence is well described when considering nuclear pair dynamics, with contributions of larger clusters giving significant corrections only when the effects of common bath are hardly visible.
In this setting we show that common bath effects should be observable in two-qubit echo signal for $d\! \leq 3$ nm, and that up to $d\! \approx \! 2$ nm (a realistic interqubit distance \cite{Jakobi2016}) pronounced non-Gaussian effects of strongly-coupled common bath should be visible for most spatial realizations of the bath. Furthermore, we show that when the environmental influence can be described as two classical noises affecting the qubits, these noises exhibit anti-correlation.

The paper is organized in the following way. In Section \ref{sec:model}
we present the qubit and bath Hamiltonian specific to the case of NV centers in diamond, and we discuss the issues related to presence of dipolar coupling between two qubits located close one to another.   
Section \ref{sec:entanglement_decay} contains a discussion of relation between two-qubit entanglement and two-qubit coherence for Bell states subjected to pure dephasing, followed by general formulation of two-qubit spin echo decoherence and the discussion of ways in which common environment affects the decoherence of $\ket{\Phi}$ and $\ket{\Psi}$ Bell states. Crucially, in Sec.~\ref{sec:correlated} we derive the relation between coherences of these states when the two qubits are subjected to classical Gaussian noise. The degree to which this relation is fulfilled when decoherence is calculated with a well-controlled quantum mechanical method (CCE in our case), is one of the main subjects of considerations presented in subsequent Sections. The two-qubit generalization of the CCE method is then described in Section \ref{sec:theory}, in which we also give analytical formulas for contributions of two-spin clusters, and discuss the behavior of these contributions for nuclei weakly and strongly coupled to the qubits. Section \ref{sec:features} contains a discussion of general features of echo decay for electron spin qubits interacting with sparse nuclear baths and convergence of CCE method in the case of NV center qubit interacting with $^{13}$C bath of natural concentration. Finally, in Section \ref{sec:results} we present and discuss our results on two-qubit coherence decay, while focusing on features brought upon by common part of environment of the two qubits.

\section{The Model}  \label{sec:model}
We first introduce the Hamiltonian that will be used to derive all the results of the paper, and then, in Section \ref{sec:gradient} we discuss the modifications of this Hamiltonian caused by addition of magnetic field gradient. The gradient is an important element of a realistic experimental setup: it allows for separate addressing of the two qubits, and extends the applicability of pure dephasing approximation to calculation of all possible two-qubit coherences. However, as we show later in the paper, the modifications of the Hamiltonian due to presence of realistic gradient large enough to achieve all the above goals, do not have a significant influence on any of the results presented in this paper. 

\subsection{Pure dephasing Hamiltonian for two qubits interacting with a nuclear bath}
The general structure of the pure dephasing Hamiltonian is given by
\begin{equation}
\HH=\sum\limits_{\alpha=1,2}\left[\Omega_{\alpha}\hat{S}^{(\alpha)}_z+\Delta_0\left(\hat{S}^{(\alpha)}_z\right)^2\right]+\HH_E+\sum\limits_{\alpha=1,2}\hat{S}^{(\alpha)}_z\hat{V}^{(\alpha)}
\label{eq:Hpd}
\end{equation}
where $\alpha$ enumerates the qubits, $\Omega_{\alpha}$ is the Zeeman splitting of qubit $\alpha$, $\Delta_{0}$ is the zero-field splitting term present for qubits based on spin $S\! > \! 1/2$ embedded in crystalline environment (such as NV centers),  $\HH_E$ is the Hamiltonian of the environment and $\hat{V}^{(\alpha)}$ is the environmental operator that couples to qubit $\alpha$.

For an NV center the quantization axis $z$ is determined by the direction of vector connecting the N impurity to nearest-neighbor vacancy \cite{Doherty_PR13}. Four such directions are possible in diamond lattice, so in general the two NV center qubits could have distinct quantization axes (as in \cite{Dolde_NP13}, were distinct axes allowed for separate addressing of the qubits). 
 However, in such a situation the fact that the two qubits, located one close to another, interact with a common bath, would be obscured, as each qubit would sense a different (rotated) nuclear bath.
Therefore, here we consider a specific case of parallel quantization $z$-axes for the NV centers, and we choose the direction of global magnetic field parallel to them, see Fig.~\ref{fig:scheme}. Note also that we consider the case in which the vector connecting the positions of the two qubits is perpendicular to the $z$ quantization axis. 
The Zeeman splittings are then given by $\Omega_0 = -\gamma_e B^0_z$ with $\gamma_e=28.02$ GHz/T and $B^0_z$ being the magnetic field along the $z$ direction. The zero-field splitting  is given by $\Delta_0=2.87$ GHz, but its value is irrelevant for subsequent calculations.

Low energy degrees of freedom of NV centers constitute a three-dimensional space, corresponding to spin $S\! =\! 1$. The qubit can be based on $m\!= \! 0$ and $m\! = \! 1$ sublevels, or on $m\! =\! \pm 1$ sublevels. We consider both possibilities, but most of the results will be given for $m\! = \! 0$, $1$ case, with $m\! =\! \pm 1$ results presented when contrasting them with the former case will be enlightening. We will label the signle-qubit states by $\ket{m}$, and the two-qubit states by $\ket{m_1,m_2}$.

The nuclear bath Hamiltonian consists of Zeeman term,
\begin{equation}
\HH_{Z}^{nuc}=\omega\sum\limits_{k} \hat{I}^k_z \,\, ,
\end{equation}
where $\omega \! =\! - \gamma_{^{13}C} B^0_z$ and $\gamma_{^{13}C}=10.71$ MHz/T is the nuclear gyromagnetic ratio for $^{13}$C, and the dipolar interaction term
\begin{equation}
\HH^{nuc}_{dip}=\sum\limits_{k<l}\,\sum\limits_{i,j=x,y,z}\hat{I}^k_i\mathbb{B}^{i,j}_{k,l}\hat{I}^l_j \,\, ,
\end{equation}
where $\mathbb{B}$ is a nuclear dipolar interaction tensor. For magnetic fields considered here, only the interactions conserving the Zeeman energy are relevant, allowing secular approximation for the dipolar couplings:
\begin{equation}
\HH^{nuc}_{dip}\approx \sum\limits_{k< l} B_{k,l}(\hat{I}^k_+\hat{I}^l_-+\hat{I}^k_-\hat{I}^l_+-4\hat{I}^k_z\hat{I}^l_z) \,\, .
\label{eq:Hnucdip}
\end{equation}
where:
\begin{equation}
B_{k,l}=\frac{\mu_0\left(\gamma_{^{13}C}\right)^2}{4\pi r_{k,l}^3}(1-3\cos^2\phi_{k,l}),
\end{equation}
with $\mu_0$ being the magnetic permeability of vacuum, $r_{k,l}$ the distance between nuclei and $\phi_{k,l}$ the angle between the vector connecting interacting nuclei and the $z$ direction of the magnetic field.

The pure dephasing interaction between qubit $\alpha$ and the nuclei is given by:
\begin{equation}
\hat{V}^{(\alpha)} = \sum\limits_{k}\sum\limits_{j=x,y,z} \hat{S}^{\alpha}_{z}\mathbb{A}^{z,j}_{\alpha,k}\hat{I}^k_j \,\, , \label{eq:Hhffull}
\end{equation}
where $\mathbb{A}$ is the hyperfine coupling tensor. Here, we consider only its dipolar part, because the Fermi contact part is very short-ranged for the NV center which is a deep defect, and we neglect the cases in which the qubit has a nuclear spin as a nearest or next-nearest neighbour.
Finally, we consider here magnetic fields $B\! \geq \! 0.3$ T, for which the influence of transverse couplings with nuclei $\propto \mathbb{A}^{z,x/y}\hat{I}_{x/y}$ is suppressed due to quick Larmor precession of nuclear spins, and only the $ \mathbb{A}^{z,z}\hat{I}^k_{z}$ couplings have to be kept in calculation of spin echo decoherence \cite{Zhao_PRB12}.  The resulting form of the coupling is then
\begin{equation}
\hat{V}^{\alpha} \approx\sum\limits_{k} \hat{S}^{\alpha}_{z} A_{\alpha,k}^{z,z} \hat{I}_{z}^k \,\, ,
\end{equation}
with 
\begin{equation}
A_{\alpha,k}^{z,z} = \frac{\mu_0\gamma_{^{13}C}\gamma_e}{4\pi R_{\alpha,k}^3}(1-3\cos^2\theta_{\alpha,k}),
\end{equation} 
where $R_{\alpha,k}$ is the distance between $k$-th nucleus and the qubit $\alpha$ and $\theta_{\alpha,k}$ is the angle between qubit energy quantization axis (the $z$ axis here) and displacement vector between nucleus and the qubit.

\subsection{Inclusion of magnetic field gradient} \label{sec:gradient}
\begin{figure}[tb]
	\centering
\includegraphics[trim={5cm 6.5cm 8cm 6cm},clip,width=\columnwidth]{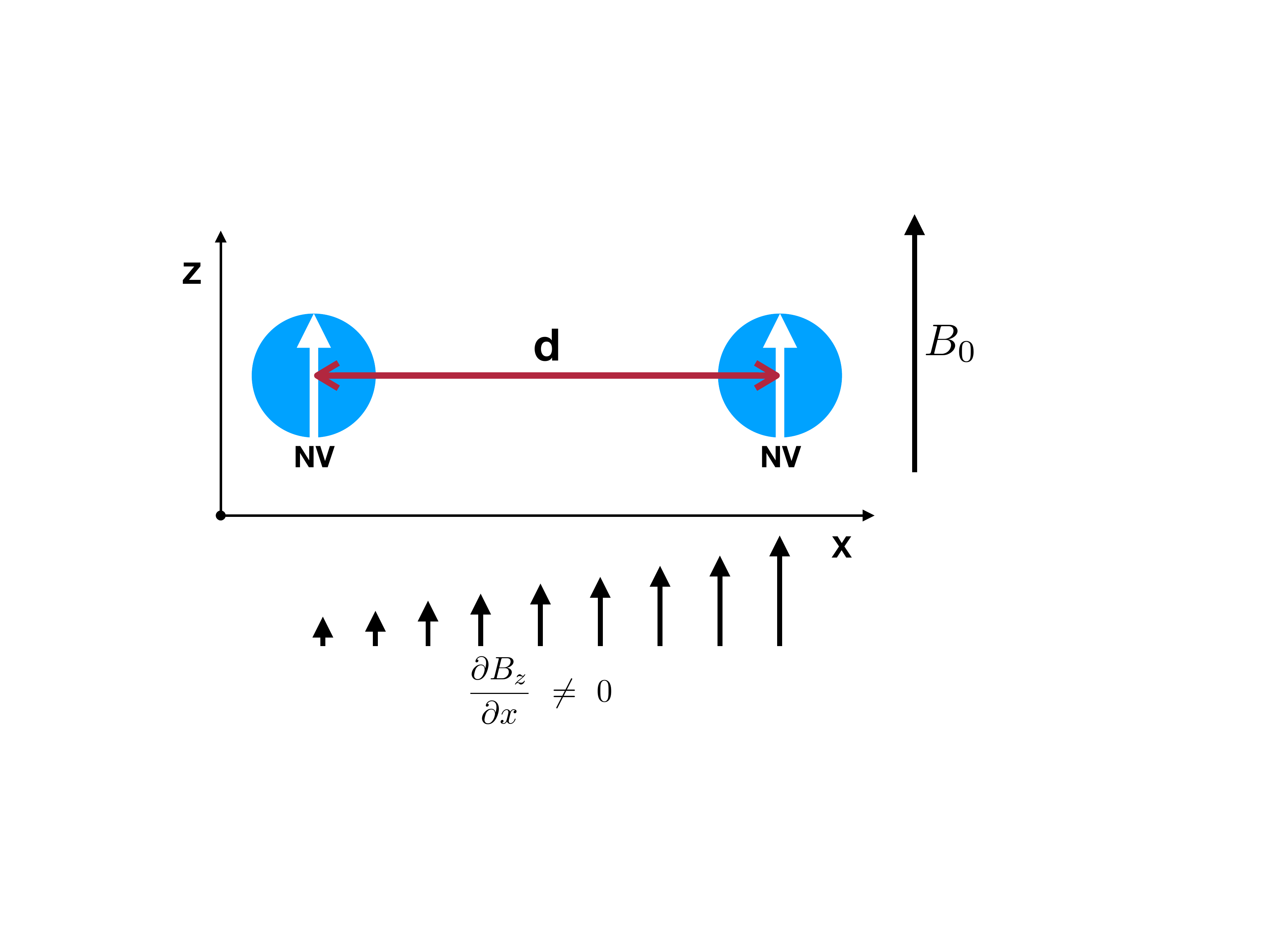}
	\caption{Physical setting: two NV centers, spatially separated by $d$, have parallel quantization axes and constant magnetic field $B_0$ is applied along the same direction. Magnetic field gradient is applied along the axis connecting both qubits. 
	}\label{fig:scheme}
\end{figure}

The Hamiltonian of a realistic system of two qubits contains also the inter-qubit dipolar interaction,
\begin{equation}
\HH_{\mathrm{dip}}^{(1,2)} = D_{12} (\hat{S}^1_{+}\hat{S}^{2}_{-} + \hat{S}^1_{+}\hat{S}^{2}_{-} - 4\hat{S}^1_{z}\hat{S}^{2}_{z} ) \,\, ,
\end{equation}
where for geometry shown in Fig.~\ref{fig:scheme} we have
$D_{12} \! = \! \mu_0\gamma_{e}^2 / 4\pi d^3$, 
with $d$ being the distance between the qubits. 

In order to consider nuclear bath induced pure dephasing of two-qubit coherences, described by the Hamiltonian given in the previous Section, the flip-flop terms in the above interaction have to be suppressed. This happens naturally when one considers NV qubits based on $m_s \! = \! \pm 1$ sublevels of spin-one manifold. Then the  $\Psi$ Bell states that are superpositions of $\ket{1,-1}$ and $\ket{-1,1}$ are coupled by $D_{12}$ to $\ket{0,0}$ state that is energetically removed by $\approx 2 \Delta_{0} \! \gg \! D_{12}$ (which is already true for $d$ smaller than the lattice constant) from them, while $\Phi$ Bell that are superpositions $\ket{1,1}$ and $\ket{-1,-1}$ are simply unaffected by the flip-flop term. For NV qubits based on $m_s=0$, $1$ sublevels, $\Phi$ states are superpositions of $\ket{00}$ and $\ket{11}$, and the first of these states is coupled by flip-flop term to $\ket{1,-1}$ and $\ket{-1,1}$ states, that are again removed by $\approx 2\Delta_{0} \! \gg \! D_{12}$ in energy. However, the $\Psi$ states are built out of $\ket{0,1}$ and $\ket{1,0}$ states that are directly coupled by the flip-flop. In order to suppress this coupling we have to introduce a  magnetic field gradient resulting in nonzero $|\Delta \Omega| \! \equiv \! |\Omega_1 - \Omega_2|$, and for $|\Delta \Omega|\! \gg \! D_{12}$ we can again neglect the presence of the flip-flop interaction. It should be noted that the presence of such a gradient is also necessary in order to separately address the two centers having parallel quantization axes, since in this case $\Omega_1 \! \neq \! \Omega_2$ is the only source of difference in energy splittings of the two qubits. 

\begin{figure}[tb]
	\centering
	\includegraphics[width=\columnwidth]{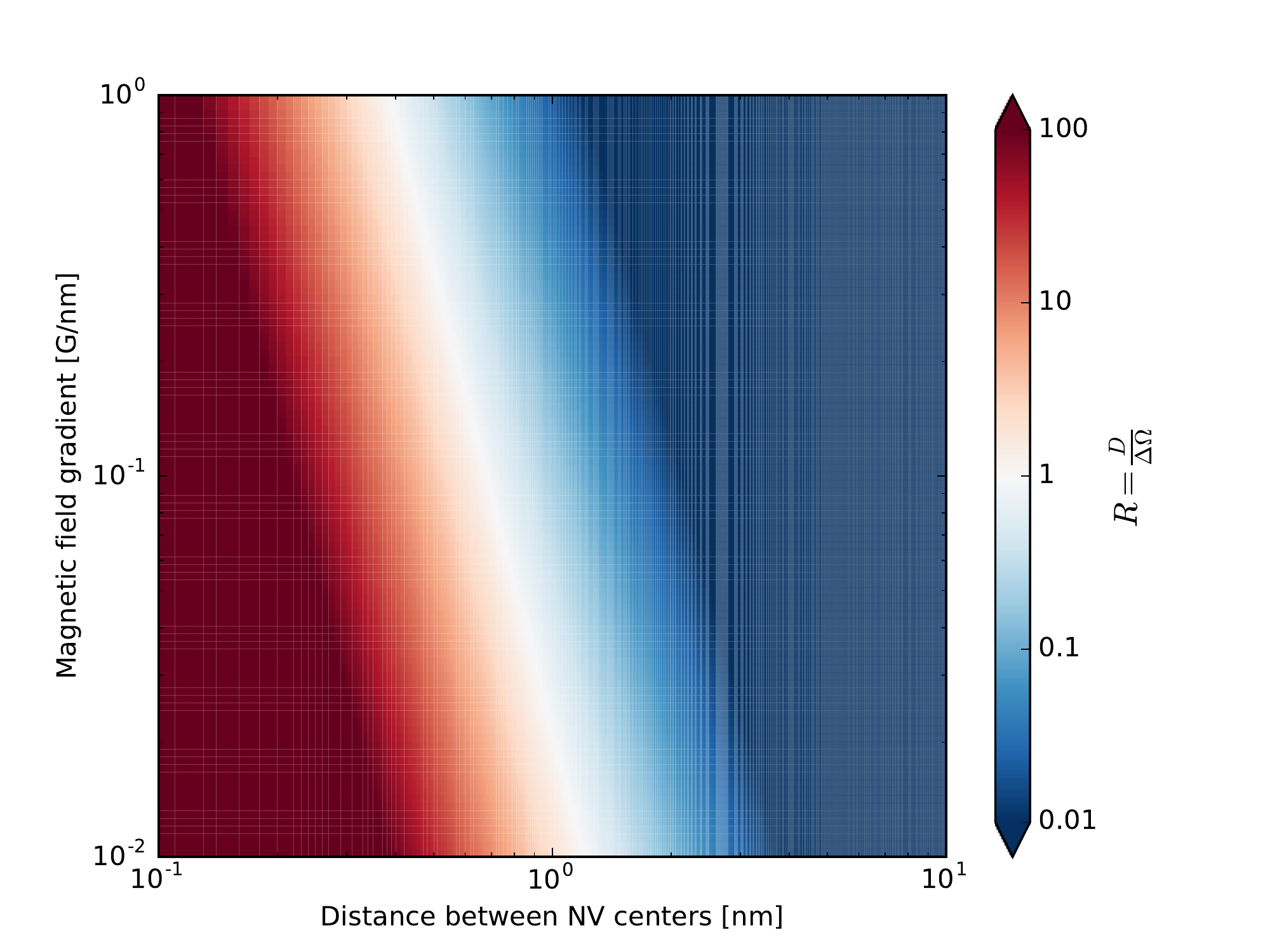}
	\caption{Two-dimensional color graph of ratio $R=D/\Delta\Omega$, where $\Delta\Omega$ is Zeeman energy mismatch between the two centers, and $D$ is their dipolar coupling as a function of magnetic field gradient and distance between NV centers. 
	}\label{fig:flipflop}
\end{figure}

In Fig.~\ref{fig:flipflop} we plot the ratio $R\! \equiv \! D_{12}/\Delta\Omega$ as a function of interqubit distance $d$ and the magnitude of magnetic field gradient along the direction of vector connecting the qubits. For $R\! \ll \! 1$ the influence of interqubit flip-flop can be safely ignored when considering bath-induced dynamics of superpositions of $\ket{0,1}$ and $\ket{1,0}$ states. For example, for $d\!=2$ nm, gradient larger than 0.05 G/nm is enough to achieve $R\!<\! 0.1$. Even 2 orders of magnitude higher magnetic field gradients are experimentally viable by putting a nanomagnet in the vicinity of the qubit, as shown e.g. in \cite{Pioro-Ladriere2008}, where it has been used for singlet-triplet qubit in GaAs double quantum dot. In the following computations we will assume a constant value of gradient $\Delta B$=1 G/nm.

The magnetic field gradient introduced above corresponds to nonzero value of $\partial B_{z}/\partial x$. In the small area containing the two qubits and their immediate surroundings we can write then $B_{z}(x) \approx B^{0}_{z} + (x-x_0)\partial B_{z}/\partial x$. Of course this results also in spatial dependence of nuclear Zeeman splittings $\omega_{k} = -\gamma_{^{13}C} B_z(x_k)$ where $x_k$ is the $x$ component of the position of $k$-th nuclear spin. Furthermore, Maxwell's equations require $\partial B_{x}/\partial z = \partial B_{z}/\partial x$, so that the magnetic field has to acquire nonzero $x$ component. We will come back to discussion of these terms in the Hamiltonian later - for now let us assert that for $B^{0}_z \! =\! 300$ mT and $\Delta B \! =\! 1$ G/nm all these corrections to nuclear Hamiltonian will have negligible influence on decoherence.

\section{Decay of entanglement due to correlated environments}  \label{sec:entanglement_decay}

\subsection{Entanglement of Bell states subjected to pure dephasing}
Let us denote the states of a single qubit by $\ket{\uparrow}$ and $\ket{\downarrow}$ - in case of NV centers these labels correspond to either $m\! =\! 0$, $1$ or $m\! = \! \pm 1$. We consider pure dephasing interaction in this basis, i.e. the case when qubit-environment interaction is diagonal in the basis of product states $
\ket{1}\!\equiv \! \ket{\uparrow\uparrow}$, $\ket{2}\! \equiv \!\ket{\uparrow\downarrow}$, $\ket{3} \! \equiv \! \ket{\downarrow\uparrow}$ and $\ket{4} \! \equiv \! \ket{\downarrow\downarrow}$. The entanglement of a mixed state resulting from subjecting an initial Bell state to such an interaction is simply proportional to the single nontrivial coherence present in the two-qubit reduced density matrix. This is a special case of a general result for so-called 'X' states (superpositions and mixtures of Bell states)\cite{Yu_QIC07,Szankowski_QIP15,Bragar_PRB15}, i.e. the concurrence \cite{Wooters_PRL98} that measures the entanglement \cite{Horodecki_RMP09,Plenio_QIC07,Aolita_RPP15} of the two-qubit state is given by $C \! = \! 2\max\left(0,|\rho_{14}|-\sqrt{\rho_{22}\rho_{33}},|\rho_{23}|-\sqrt{\rho_{11}\rho_{44}}\right)$. In case of Bell states subjected to pure dephasing we have then $C\!= \! 2|\rho_{14}|$ for $\ket{\Phi}$ states and  $C\!= \! 2|\rho_{23}|$ for $\ket{\Psi}$ states.
From here on, we discuss the time dependence of the non-zero coherence for a given initial Bell state. For example $\ket{\Psi}$ state built out of $\ket{0}$ and $\ket{1}$ states of the NV centers (i.e. for $\ket{\Psi^{\pm}_{01}}=\frac{1}{\sqrt{2}}\left(\ket{01}\pm\ket{10}\right)$ we have the \textit{decoherence function} at time $t$ given by:
\begin{equation}
W^{\Psi}_{01}(T)\equiv\frac{\rho_{01,10}(T)}{\rho_{01,10}(0)} \,\, .\label{eq:WPsi}
\end{equation}
Decoherence function defined in this way is precisely equal to the concurrence of the decohered state, since $\rho_{01,10}(0)=\pm 1/2$. Similarly, for $\ket{\Phi^{\pm}_{01}}=\frac{1}{\sqrt{2}}\left(\ket{00}\pm\ket{11}\right)$ state we have:
\begin{equation}
W^{\Phi}_{01}(T)\equiv \frac{\rho_{00,11}(T)}{\rho_{00,11}(0)} \,\, . \label{eq:WPhi}
\end{equation}
The entanglement of the Bell states built out of $m\! =\! \pm 1$ states of the NV centers is determined by analogously defined coherences $W^{\Psi}_{1-1} (T)$ and $W^{\Phi}_{1-1}(T)$.

It should be noted that nonzero coherences of these kinds are present also when a separable state of two qubits is prepared, e.g.~for initial maximal superposition state, $\prod_{q=1,2}\frac{1}{\sqrt{2}}(\ket{0}_{q}+\ket{1}_{q})$, we have both $\rho_{01,10}(0)$ and $\rho_{00,11}(0)$ equal to $1/4$, and the decay of both $\Phi$-type and $\Psi$-type coherences can be observed. However, the amplitude of the signal will be lower than for appropriate Bell states. Note that the decoherence functions defined above and used in the following are normalized by the initial value of coherence, and one should keep the above practical consideration in mind when choosing between dealing with entangled states (creation of which requires additional effort), or measuring two-qubit coherence on separable states (which are easy to prepare, but give less signal). 

\subsection{Two-qubit coherence echo decay}  \label{sec:general_echo} 
We assume here that the system is initialized in a product state of the qubits and the environment, given by $\ket{\phi}\bra{\phi}\otimes\hat{\rho}_E$ with $\ket{\phi}$ being one of the Bell states of the pair of qubits and $\hat{\rho}_E$ is the density operator of the environment, that we take here to be maximally mixed, $\hat{\rho}_E \! \propto \! \mathbb{\one}$, as it is appropriate for nuclear baths at room, or even cryogenic, temperatures.

We consider a two-qubit generalization of echo protocol, in which each of the qubits is subjected to a $\pi$-rotation (assumed to be instantaneous) about one of in-plane axes, at time $T/2$. Then the coherence is read out at time $T$. In the following, we will also assume that the readout is preceded by an additional $\pi$ pulse, for the purpose of definition of the $W$ functions as in Eqs.~(\ref{eq:WPsi}-\ref{eq:WPhi}). The resulting expressions are then:
\begin{align}
W^{\Psi}_{ab}(T)=\Tr_E\left(\hat{\rho}_Ee^{i\hat{H}_{ba}\frac{T}{2}}e^{i\hat{H}_{ab}\frac{T}{2}}e^{-i\hat{H}_{ba}\frac{T}{2}}e^{-i\hat{H}_{ab}\frac{T}{2}}\right) \,\, , \label{eq:WPsiecho}\\
W^{\Phi}_{ab}(T)=\Tr_E\left(\hat{\rho}_Ee^{i\hat{H}_{bb}\frac{T}{2}}e^{i\hat{H}_{aa}\frac{T}{2}}e^{-i\hat{H}_{bb}\frac{T}{2}}e^{-i\hat{H}_{aa}\frac{T}{2}}\right) \,\, , \label{eq:WPhiecho}
\end{align}
where $\HH_{ab}$ are given by
\begin{align}
\HH_{ab} & =a\hat{V}^{(1)}+b\hat{V}^{(2)}+\HH_E \,\, , \\
& = \frac{a+b}{2}(\hat{V}^{(1)}+\hat{V}^{(2)})+\frac{a-b}{2}(\hat{V}^{(1)}-\hat{V}^{(2)})  +\HH_E \,\, . \label{eq:Hab}
\end{align}

\subsection{Effects of common environment on two-qubit dephasing} \label{sec:correlated}
Let us look at how we can quantify the effects of common environment on two-qubit dephasing. A natural point of reference is dephasing caused by two completely uncorrelated environments. Let us define single-qubit coherence for qubit $\alpha$ as
\begin{equation}
w^{(\alpha)}_{ab}(T)=\Tr_{E}\left(\hat{\rho}_{E}e^{iH_a\frac{T}{2}}e^{iH_b\frac{T}{2}}e^{-iH_a\frac{T}{2}}e^{-iH_b\frac{T}{2}}\right) \label{eq:Wsingleecho}
\end{equation}
with $H_{a}=a\hat{V}^{(\alpha)}+\hat{H}_{E}$. This quantity corresponds to normalized single-qubit coherence $\rho^{(\alpha)}_{ab}(T)$ obtained when the second qubit is physically absent or decoupled from the bath. The latter situation occurs when the second qubit is initialized in $\ket{0}$ state -- then the $w^{(\alpha)}_{ab}$ quantity corresponds to normalized two-qubit coherences $\rho_{a0,b0}$ or $\rho_{0a,0b}$, depending on $\alpha\! =\! 1$, $2$.

For interqubit distance $d$ larger than a certain value, it should be possible to approximate the two-qubit dephasing by assuming that each qubit interacts with a separate environment. If in qubit-bath interaction $\hat{V}^{(\alpha)}$ we neglect $A^{z,z}_{\alpha,k}$ couplings smaller than a certain threshold value, then due to the fact that $A^{z,z}_{\alpha,k}$ decreases with qubit-nucleus distance, beyond some $d$ we will have $[\hat{V}^{(1)},\hat{V}^{(2)}] \! = \! 0$. If we then also neglect dipolar interactions between nuclei belonging to environments of distinct qubits, and add the assumption of no inter-nuclear correlations in the density matrix of the bath (obviously valid in the considered here case of high temperature bath with $\hat{\rho}_{E} \! \propto \! \mathds{1}$), then 
the averages in Eqs.~\eqref{eq:WPsiecho} and \eqref{eq:WPhiecho} factorize into products of separate averages over environments of qubits $\alpha\! = \! 1,2$:
\begin{equation}
W^{\Psi}_{ab}(T)=W^{\Phi}_{ab}(T)=w^{(1)}_{ab}(T)w^{(2)}_{ab}(T) \,\, .
\end{equation}
The first question that we want to answer in this paper is at what $d$ we can expect noticeable deviations from this relation in the case of NV center spin qubits interacting with $^{13}$C bath of natural concentration. The second question concerns the quantitative nature of these deviations. 

Since the difference between the two-qubit coherences $W^{\Psi/\Phi}(T)$ and the product of two appropriate single-qubit coherences is the basic signature of the common environment affecting the qubits, we introduce the following $\Lambda(T)$ function 
\begin{equation}
\Lambda^{\Psi/\Phi}_{ab}(T) \equiv \frac{W^{\Psi/\Phi}_{ab}(T)}{w^{(1)}_{ab}(t)w^{(2)}_{ab}(T)} \,\, .
\label{eq:Lambda} 
\end{equation}
The effects of common environment on two-qubit decoherence are visible whenever $\Lambda(T)$ appreciably differs from unity.

When the two qubits are coupled in the same way to the common environment, i.e.~when they are exposed to {\it perfectly correlated environmental noises}, we have $\hat{V}^{(1)}=\hat{V}^{(2)}$. From Eq.~(\ref{eq:Hab}) we see that $\hat{H}_{ab}\! =\! \hat{H}_{ba}$ then, and according to Eq.~(\ref{eq:WPsiecho})  we have $W^{\Psi}_{ab} \!= \!1$, so that  $\ket{\Psi_{ab}}$ states do not decohere, and $\Lambda^{\Psi}_{ab}(T) \! =\! 1/w^{(1)}_{ab}(T)w^{(2)}_{ab}(T)$ grows to $\infty$ as the single-qubit coherences decay towards zero with increasing $T$.
On the other hand, for \textit{perfectly anti-correlated} influence of $E$ on both qubits, $\hat{V}^{(1)}\! = \!-\hat{V}^{(2)}$, the $\ket{\Phi_{ab}}$ states do not decohere. These Bell states belong to decoherence-free subspaces \cite{Lidar_ACP14} for the respective highly symmetric couplings between the qubits and $E$. 

Another case in which we can make general statements on behavior of $\Lambda^{\Psi/\Phi}_{ab}(T)$ functions is when $\hat{V}^{(\alpha)}$ couplings are replaced with classical and Gaussian stochastic processes $\xi_{\alpha}(t)$. In other words, the influence of $E$ on the qubits is assumed then to be well-described as action of classical Gaussian noise. Tracing over $E$ in expressions for coherences is replaced by averaging over all the realizations of noises $\xi^{(\alpha)}(t)$, the statistical properties of which reflect the behavior of the bath:
\begin{equation}
W^{\Psi/\Phi}_{ab}(t)={\bigg\langle} e^{-i(a-b)\int\limits_{0}^{T}\left[\xi_1(t)\mp\xi_2(t)\right]f_T(t)\text{d}t}{\bigg\rangle}
\end{equation}
where $f_T(t)$ is the time-domain \textit{filter function} \cite{deSousa_TAP09,Szankowski_JPCM17} encoding the sequence of pulses applied to the qubits, with 
$f_T(t)=\Theta(t)\Theta(\frac{T}{2}-t)-\Theta(t-\frac{T}{2})\Theta(T-t)$ for two-qubit echo considered here.  
The averaging over noise realizations can be easily done due to the assumption of Gaussian noise statistics, giving \cite{Szankowski_PRA16,Szankowski_JPCM17} 
\begin{align}
W^{\Psi/\Phi}_{ab}(T)&=e^{-(a-b)^2[\chi_{11}(T)+\chi_{22}(T)\pm 2\chi_{12}(T)]},\nonumber\\
&=w^{(1)}_{ab}(T)w^{(2)}_{ab}(T)e^{\pm 2(a-b)^2\chi_{12}(T)} \,\, , \label{eq:Wchi12}
\end{align}
in which the $\chi_{\alpha\beta}(T)$ are the \textit{attenuation functions}. For the considered here case of both qubits exposed to the same echo pulse sequence 
they are given by
\begin{align}
\chi_{\alpha\alpha}(T)&=\frac{1}{2}\int\limits_{-\infty}^{\infty}S_{\alpha\alpha}(\omega)|\tilde{f}\,^{T}(\omega)|^2\frac{\text{d}\omega}{2\pi}\\
\chi_{12}(T)&=\frac{1}{2}\int\limits_{-\infty}^{\infty}S^R_{12}(\omega)|\tilde{f}\,^{T}(\omega)|^2\frac{\text{d}\omega}{2\pi} \,\ ,
\end{align}
where $S_{\alpha\alpha}(\omega)$ is the spectrum of $\xi^{(\alpha)}(t)$ noise, the Fourier transform of its autocorrelation function $\langle\xi_{\alpha}(t)\xi_{\alpha}(0)\rangle$, while $S^R_{12}(\omega)$ is the real part of the cross-spectrum of the two noises, being a Fourier transform of cross-correlation function $\langle\xi_{1}(t)\xi_{2}(0)\rangle$ \cite{Szankowski_PRA16}.

In this setting, the correlations between the noises, experienced by the qubits, are all contained in $e^{\pm 2(a-b)^2\chi_{12}(T)}$ term in Eq. (\ref{eq:Wchi12}). This means that the $\Lambda(T)$ functions defined in Eq. \eqref{eq:Lambda} fulfill 
\begin{equation}
\Lambda^{\Psi}_{ab}(T) =  \frac{1}{\Lambda^{\Phi}_{ab}(T)} \,\, . 
\label{eq:GaussianLambda}
\end{equation}
This is an important result specific to the Gaussian noise approximation. Observation of breaking of this relationship by two-qubit decoherence proves that the influence of the environment {\it cannot} be modeled by treating it as classical Gaussian noise. 
It has to be stressed now that the Gaussian noise approximation is often used because it leads to simple analytical formulas connecting the attenuation functions with spectral densities that have a simple physical interpretation. However, its applicability to the case of rather small environments is often questionable - but methods for recognizing that this approximation fails from a few simple measurements on a qubit (or qubits) are lacking \footnote{One can check if the noise is Gaussian by performing a procedure of quantitative characterization of non-Gaussian features of noise, and assuming that the noise is indeed Gaussian if this procedure gives a null result. However, recently proposed methods of such characterization employing a single qubit \cite{Norris_PRL16} require measurement of coherence under influence of many distinct pulse sequences, and further nontrivial analysis of the obtained results.}. 

In this work we perform quantum-mechanical calculation of two-qubit dephasing caused by an environment of interacting nuclei. With exception of one case (that of two qubits positioned very close one to another), neither of the above-discussed cases applies to the system that we consider: we have $\hat{V}^{(1)}\! \neq \! \hat{V}^{(2)}$ and the applicability of Gaussian noise approximation needs to be ascertained by comparing its predictions with results of CCE calculations. As we discussed in the Introduction, we expect the Gaussian noise model to work well when the relevant environment consists of many entities that are weakly coupled to the qubit(s) and are approximately uncorrelated - but this situation is by no means obvious when dealing with a sparse nuclear bath.

We propose here to use the breaking of relationship from Eq.~(\ref{eq:GaussianLambda}) as a ``witness'' of non-Gaussianity of environmental noise experienced by the qubits coupled to a common environment. When the qubits are close enough one to another to assume that they are both exposed to the same environmental influence (i.e.~$\hat{V}^{(1)} \! =\! \hat{V}^{(2)}$), observing that Eq.~(\ref{eq:GaussianLambda}) is fulfilled, means that we can model the environmental influence by replacing $\hat{V}^{(\alpha)}$ with a common Gaussian noise $\xi(t)$. For more distant qubits, in the same way we can make a statement on the nature of the noise that originates from the part of the environment that visibly affects both of the qubits and leads to nonzero $\chi_{12}(T)$ function. 

Note that the ``non-Gaussianity witness'' proposed here has a physical origin similar to the so-called ``anomalous decoherence'' effect described in \cite{Zhao_PRL11,Huang_NC11}. In the Gaussian noise model, the attenuation functions $\chi$ depend quadratically on qubit-environment couplings - note that all the $\chi_{\alpha\beta}$ are multiplied by $(a-b)^2$ in Eq.~(\ref{eq:Wchi12}). This scaling leads to a relation between cases of $a=0,\,b=1$, and $a=-1,\,b=1$ (so-called ``double coherences'') that in a single-qubit case reads
\begin{equation}
w^{(1)}_{1-1}(T) = \left[w^{(1)}_{01}(T)\right]^4 \,\, ,  \label{eq:ADE}
\end{equation}
and the same holds for two-qubit coherences $W^{\Psi/\Phi}_{1-1}$ and $W^{\Psi/\Phi}_{01}$. In \cite{Zhao_PRL11,Huang_NC11} it was noted that microscopic treatment of a quantum bath (using the CCE method) gives results that often disagree with this relation, showing thus that there are parameter regimes in which the influence of nuclei on an NV center spin qubit {\it cannot} be modeled as classical Gaussian noise. Tuning between Gaussian and non-Gaussian regimes with magnetic field was shown in \cite{Reinhard_PRL12}, in which these regimes were referred to as ``classical'' and ``quantum''. Equation (\ref{eq:GaussianLambda} has the same origin as Eq.~(\ref{eq:ADE}), but its violation gives information pertaining not to the total noise experienced by each qubit, but on to the  noise originating from the part of environment that is appreciably coupled to both the qubits.

\section{Calculation of two-qubit decoherence}  \label{sec:theory}
\subsection{Cluster-correlation expansion for two-qubit decoherence in echo experiment} \label{sec:twoqubitCCE}
In the CCE method the dephasing of the qubit interacting with a large interacting bath is approximated by calculating contributions to dephasing coming from smaller parts of the bath - clusters of nuclei - while taking into account only intra-cluster interactions between the nuclei. 
CCE is self-consistent in a sense that if a system contains $M$ nuclei and we calculate decoherence up to CCE-$M$ order, we end up with an exact result. For a Hamiltonian consisting of Zeeman splittings, spin-flipping terms, one has to start from calculation of coherence for each single bath spin. We denote contributions from clusters to qubit(s) decoherence as $\mathcal{L}_{\mathbb{C}}$, where $\mathbb{C}$ is a cluster of a given size. $\mathcal{L}_{\mathbb{C}}$ is simply the decoherence function for a qubit (or, as is the case here, a pair of qubits) calculated by keeping only the spins from cluster $\mathbb{C}$ in environment-qubit(s) interaction and the Hamiltonian of the environment. Then the CCE-1 approximation to decoherence is a product of single-nucleus contributions:
\begin{equation}
W^{(\text{CCE-}1)}=\prod\limits_{n=1}^{N} \mathcal{L}_n(t)
\end{equation}
where $n$ enumerates nuclei that form the bath and $N$ is the total number of those. Each $\mathcal{L}_n(t)$ for echo in case of both single- and two-qubit coherences, can be calculated according to Eqs. (\ref{eq:WPsiecho}--\ref{eq:Wsingleecho}), with Hamiltonian of the environment consisting only of the Zeeman splitting of considered nucleus. When transverse couplings to the qubit are negligible, as in our case, when the whole system is in magnetic field $B>$100 mT, non-interacting bath of nuclei does not give any contribution to decoherence, i.e., $\mathcal{L}_n(t)=1$. 

Second order contribution, CCE-2, is a product of {\it irreducible} contributions to decoherence of two-spin clusters, i.e.~nuclear pairs. Technically, we calculate coherence of a qubit interacting with each pair and divide the result by contributions of each nucleus in the given pair:
\begin{equation}
W^{(\text{CCE-}2)}=W^{(\text{CCE-1})}(t)\cdot\prod\limits_{(k,l)} \frac{\mathcal{L}_{kl}(t)}{\mathcal{L}_k(t)\mathcal{L}_l(t)}
\end{equation}
where $k,\,l$ enumerates nuclear spins. For the calculation of each $\mathcal{L}_{kl}(t)$, we need to consider qubit(s) interacting with a pair of nuclei $k$ and $l$ and now the Hamiltonian of the environment, as in procedure described in Eqs. (\ref{eq:WPsiecho}--\ref{eq:Wsingleecho}) contains not only Zeeman splittings of each of the nuclei, but also coupling between them. In this paper, we shall use the dipolar interaction as described in Eq. \eqref{eq:Hnucdip}.

Using the same logic, we can calculate coherence up to arbitrary CCE-$M$:
\begin{equation}
W^{{\text{CCE-}M}}= W^{\text{CCE-}(M-1)}\cdot\prod_{M\text{-cl.}}\widetilde{\mathcal{L}}_{(M-\text{cl.})}
\end{equation} 
where $\widetilde{\mathcal{L}}_{(M-\text{cl.})}$ is an irreducible contribution from unique cluster of $M$ nuclei - $\mathcal{L}_{M}$ - divided by products of decoherence contributions from all the smaller clusters formed of these spins. 

\setlength{\arrayrulewidth}{0.5mm}
\setlength{\tabcolsep}{18pt}
\renewcommand{\arraystretch}{1.0}
\begin{table}[!tb]
	\centering
	\label{tab:A}
	\begin{tabular}{l|l|l}
		
		& $\mathcal{A}_+$           & $\mathcal{A}_-$\\ \hline
		$\ket{\Psi_{01}}$                                 & $\Delta A_1$              & $\Delta A_2$\\
		$\ket{\Phi_{01}}$                                 & 0                         & $\Delta A_1+\Delta A_2$\\
		$\ket{\Psi_{1-1}}$                                & $\Delta A_1-\Delta A_2$   & $-(\Delta A_1-\Delta A_2)$\\
		$\ket{\Phi_{1-1}}$                                & $(\Delta A_1+\Delta A_2)$ & $-(\Delta A_1+\Delta A_2)$\\
		$\ket{x^+_{01}}$  & $\Delta A_1$                & 0\\
		$\ket{x^+_{1-1}}$ & $\Delta A_1$                & $-\Delta A_1$             
	\end{tabular}
		\caption{Effective hyperfine couplings for a nuclear pair (with two nuclei labeled with $k$ and $l$) interacting with qubit(s). Left column contains initial qubit(s) state, with single-qubit states (of qubit $1$) given by $\ket{x^+_{01}}=\frac{1}{\sqrt{2}}\left(\ket{0}+\ket{1}\right)$ and $\ket{x^+_{1-1}}=\frac{1}{\sqrt{2}}\left(\ket{1}+\ket{-1}\right)$. $\mathcal{A}_{+}$ and $\mathcal{A}_{-}$ are, respectively, couplings entering Eq.~(\ref{eq:Lkl}) conditioned on the two states of the qubit(s) present in the superposition state from the first column. $\Delta A_{\alpha} = A^{z,z}_{q,k}-A^{z,z}_{q,l}$ is the difference of couplings of qubit $\alpha\! = \! 1$, $2$ with the two nuclei.}
\end{table}

We consider a nuclear bath of uniformly distributed $^{13}$C nuclei at room temperature. As thermal energy associated with this temperature is significantly higher than the nuclear spin Zeeman energy (e.g. magnetic field of 100 mT corresponds to 26 $\mu$K and inter-nuclear dipolar interactions  for nearest neighbour distance in diamond lattice - 0.15 nm - correspond to 9.2 $\mu$K), the bath density matrix is completely mixed, i.e., $\hat{\rho}_E \! \propto \! \hat{\one}$.
In high magnetic fields (in case of NV center in diamond, that is $B>100$ mT), the single-spin contributions to echo decay (i.e.~CCE-1 contributions), are essentially absent - the dynamical contributions from transverse $A^{z,x/y}\hat{I}_{x/y}$ terms are vanishing due to very fast Larmor precesssion of nuclear spins, while the quasi-static contributions from averaging over single-spin $A^{z,z}\hat{I}_{z}$ terms are removed by echo procedure. 
At these fields, at which we can also safely use the secular approximation of dipolar interactions between nuclei in the bath, the pairwise contribution to decoherence (CCE-2 level of approximation) can be analytically calculated for both single-qubit and Bell state coherences.  
Contribution to decoherence of both two- and single-qubit state, within the magnetic field regime justifying the above described assumptions,can be generally expressed the following way:
\begin{align}
&\mathcal{L}_{kl}=1-(\mathcal{A}_+-\mathcal{A}_-)^2b^2\times\nonumber\\
&\frac{\sin^2\left(\frac{T}{2}\sqrt{\left(\Delta\omega+\mathcal{A}_+\right)^2+b^2}\right)}{b^2+\left(\mathcal{A}_{+} +\Delta\omega\right)^2}\frac{\sin^2\left(\frac{T}{2}\sqrt{\left(\Delta\omega+\mathcal{A}_{-}\right)^2+b^2}\right)}{b^2+\left(\mathcal{A}_-+\Delta\omega\right)^2} \,\, , \label{eq:Lkl}
\end{align}
where the effective couplings $\mathcal{A}_{\pm}$ for various two-qubit and single-qubit states are given in Table \ref{tab:A}, $\Delta\omega \! \equiv \! \omega_k - \omega_l$ is the difference of Zeeman splittings of the nuclei and $b \! \equiv \! B_{kl}$ is their mutual dipolar interaction strength.

In the regime of magnetic fields considered in this paper, coherence calculated up to CCE-2 level (with CCE-1 contributions being negligible) is simply a product of the above $\mathcal{L}_{kl}$ contributions calculated for all the nuclear pairs (obviously only pairs of nuclei quite close one to another contribute significantly, as $\mathcal{L}_{kl} \! \propto \! b^2/\mathcal{A}^2$ for $b \! \ll \! \mathcal{A}$). Since the definition of $\Lambda$ functions involves a ratio of two-qubit and single-qubit coherences, at CCE-2 level of calculation , each $\Lambda(T)$ can be represented as a product of contribution of all possible $(k,l)$ pairs of nuclei:
\beq
\Lambda_{a/b}^{\Psi/\Phi}(T) = \frac{ W_{a/b}^{\Psi/\Phi}(T)}{w^{(1)}_{ab}(T)w^{(2)}_{ab}}(T) = \prod_{k,l} \Lambda^{\Psi/\Phi}_{a/b; k,l}(T) \,\, .
\eeq
	Let us look now more closely at $\Lambda^{\Psi/\Phi}_{a/b; k,l}(T)$, the contributions of individual pair of spins to $\Lambda(T)$ function characterizing the correlations in environmental noise experienced by the qubits.  

\subsection{Nuclear pair contributions to decoherence}  \label{sec:strongweak}
We focus now on contribution of single nuclear pair $(k,l)$ to decay of various two- and single-qubit coherences, given by Eq.~(\ref{eq:Lkl}) with an appropriate choice of $\mathcal{A}_{\pm}$ couplings from Table \ref{tab:A}. For clarity we will first focus on $\Delta \omega \! = \!0 $ case and on qubits based on $m\! =\! 0$, $1$ states of NV center, thus $a \! =\! 0$ and $b\! =\! 1$, and we suppress this label below. Furthermore, $\mathcal{L}_{kl}$ is  close to unity for almost all the nuclear pairs, so it is convenient to define
\begin{equation}
\mathcal{L}_{kl}(T) = 1-\delta \mathcal{L}_{kl}(T) \,\, ,
\end{equation}
and assume that $\delta \mathcal{L}_{kl} \!\ll \! 1$. The pair contribution to $\Lambda$ for a given entangled state is
\begin{equation}
\Lambda^{\Psi/\Phi}_{k,l}=\frac{1-\delta \mathcal{L}^{\Psi/\Phi}_{k,l}}{(1-\delta \mathcal{L}^{(1)}_{k,l})(1-\delta \mathcal{L}^{(2)}_{k,l})} \,\, ,
\end{equation}
and to first order in small quantities $\delta \mathcal{L}$ it is given by
\beq
\Lambda^{\Psi/\Phi}_{k,l} \approx 1-\delta \mathcal{L}^{\Psi/\Phi}_{k,l} + \delta \mathcal{L}^{(1)}_{k,l} + \delta \mathcal{L}^{(2)}_{k,l} \,\, . \label{eq:Lapp}
\eeq

In the following it will be crucial to distingiush between nuclear pairs strongly and weakly coupled to each of the qubits. We define the pairs strongly (weakly) coupled to qubit $q$ as fulfilling
\beq
\Delta A_{q} \gg b  \,\,\,  (\Delta A_{q} \ll b) \,\, .
\eeq 
This means that for strongly coupled pair, its interaction $\Delta A_{q}$ with the qubit $q$ is stronger that the intra-pair coupling $b$, so that the pair-qubit coupling has a strong effect on the pair dynamics. The opposite holds for weakly coupled pairs, which evolve primarily due to the mutual interaction of the nuclei, with the nuclei-qubit interaction playing a negligible role. It is thus intuitive that the classical noise approach to decoherence, in which the bath is a source of noisy signal affecting the qubit, but in itself {\it unaffected} by the presence of the qubit, can only be applicable when dealing with weakly coupled nuclear clusters.

When the nuclear pair $(k,l)$ is weakly coupled to both qubits, then after making an additional simplyfying assumption of $t\ll \frac{b}{\Delta A_1^2}$, $\frac{b}{\Delta A_2^2}$ we obtain
\begin{equation}
\delta \mathcal{L}^{\Psi/\Phi}_{w-w} \approx  \frac{1}{b^2}(\Delta A_1 \mp \Delta A_2)^2\sin^4\left(\frac{bt}{2}\right) \,\, ,
\end{equation}
where the subscript $w-w$ denotes the case of weak coupling to both qubits. Then, using Eq.~(\ref{eq:Lkl}) and Table \ref{tab:A} we obtain for contributions of this pair to the measure of correlation in decoherence:
\begin{equation}
\Lambda^{\Phi/\Psi}_{w-w}\approx  1 \pm \frac{\Delta A_1\Delta A_2}{b^2}\sin^4\left(\frac{bt}{2}\right) 
\,\, ,
\end{equation}
which means that $\delta \mathcal{L}_{kl}^{\Psi/\Phi} \! \approx \! \mp \delta \chi_{k,l}$.
This result is consistent with treating weakly coupled nuclear pairs effectively as a source of Gaussian noise acting on the two qubits. The contribution of large number of these pairs to decoherence is the
\begin{align}
\prod_{k,l}\Lambda^{\Psi/\Phi}_{k,l} & = \prod_{k,l} (1-\mathcal{L}_{kl}^{\Psi/\Phi}) \approx  \prod_{k,l}(1 \pm \delta \chi_{k,l}) \nonumber\\ 
& \approx e^{\pm \sum_{k,l}\delta \chi_{k,l} } \,\, ,
\end{align}
and in this case the relation $\Lambda^{\Psi} \! =\! 1/\Lambda^{\Phi}$ is fulfilled.

When a nuclear pair is strongly ($s$) coupled to qubit $1$ (i.e.~$\frac{b}{\Delta A_1}\ll 1$) and weakly ($w$) coupled to qubit $1$ (i.e.~$\frac{\Delta A_2}{b}\ll 1$), for $\frac{b^2t}{\Delta A_1 }\ll 1$ and $\frac{\Delta A_2^2t}{b}\ll 1$ the contributions to single-qubit coherences are simply:
\begin{align}
\delta\mathcal{L}^{(1)}_{s}& \approx \sin^2\left(\frac{bt}{2}\right)\sin^2\left(\frac{\Delta A_1 t}{2}\right) \,\, ,  \label{eq:strongsingleW}\\
\delta\mathcal{L}^{(2)}_{w}&\approx \frac{\Delta A_2^2}{b^2}\sin^4\left(\frac{bt}{2}\right)  \,\, , \label{eq:weaksingleW}
\end{align}
while the contributions to two-qubit coherences obtained using the same approximations are {\it both} equal to $\delta\mathcal{L}^{(1)}_{s}$ from Eq.~(\ref{eq:strongsingleW}). The main feature of the coherent of both entangled states (and the coherence of qubit $1$) is the oscillation (``the fingerprint'' of a strongly coupled pair) with high frequency $\Delta A_{1}$. However, according to Eq.~(\ref{eq:Lapp}) this oscillation is removed from the $\Lambda$ quantity measuring the effects of the common part of the bath:
\begin{equation}
\Lambda^{\Psi}_{s-w} \approx \Lambda^{\Phi}_{s-w}  \approx  1 + \delta\mathcal{L}^{(2)}_{w}  \,\, ,
\end{equation}
with $\delta\mathcal{L}^{(2)}_{w}$ given by Eq.~(\ref{eq:weaksingleW}). We see then that when a nuclear pair is strongly coupled to one qubit, and weakly coupled to the other, only the weak-coupling contribution remains in the correlation signal -- but this contribution is {\it the same} for $\ket{\Psi}$ and $\ket{\Phi}$ states, in contradition to the result expected for contribution of Gaussian noise.

Finally, let us look at the case of a nuclear pair strongly coupled to both qubits. Assuming $b^2 t/\Delta A_{1,2} \! \ll \! 1$, the single-qubit contributions can be approximated as in Eq. \eqref{eq:strongsingleW} for both qubits, but the entangled pair coherences look approximately as follows:
\begin{align}
\delta \mathcal{L}^{\Psi}_{s-s}&\approx\frac{(\Delta A_1-\Delta A_2)^2b^2}{\Delta A_1^2\Delta A_2^2}\sin^2\left(\frac{\Delta A_1 t}{2}\right)\sin^2\left(\frac{\Delta A_2 t}{2}\right) \,\ , \\
\delta \mathcal{L}^{\Phi}_{s-s}&\approx \sin^2\left(\frac{bt}{2}\right) \sin^2\left(\frac{\Delta A_1+\Delta A_2}{2}t\right) \,\, .
\end{align}
It is interesting to note that while $\delta \mathcal{L}^{\Phi}_{s-s}(t)$ is of order unity at long times, $\delta \mathcal{L}^{\Psi}_{s-s}(t)$ is on the other hand, of the order of $b^2/\Delta A^2 \! \ll \! 1$. The $\ket{\Psi_{01}}$ state is thus quite robust to depashing due to interaction with a nuclear pair that is strongly coupled to both of the qubits. 
This is in contrast to the case of strong-weak coupling in which, as noted below Eq.~(\ref{eq:weaksingleW}) both $\delta \mathcal{L}^{\Psi}$ and $\delta \mathcal{L}^{\Phi}$ are of order unity at long times.  For pair contribution to correlation we obtain, dropping terms $\sim b^2/\Delta A^2$,
\begin{align}
\Lambda^{\Psi}_{s-s} & \approx 1+  \sin^2\left(\frac{bt}{2}\right) \Big[ \sin^2\left(\frac{\Delta A_1 t}{2}\right) + \sin^2\left(\frac{\Delta A_2 t}{2}\right) \Big] \,\, , \label{eq:LPsi01} \\
\Lambda^{\Phi}_{s-s} & \approx 1+\sin^2\left(\frac{bt}{2}\right) \Big[ \sin^2\left(\frac{\Delta A_1 t}{2}\right) + \sin^2\left(\frac{\Delta A_2 t}{2}\right) \nonumber\\
&  - \sin^2\left(\frac{\Delta A_1+\Delta A_2}{2}t\right) \Big ] 
\end{align}
The results disagree (rather unsurprisingly at this point) with the Gaussian noise predictions (i.e.~$\Lambda^{\Psi/\Phi}_{k,l} \! \approx 1 \pm \delta \Lambda_{k,l}$ on the level of single-pair contributions). Note that in this case all the coherence signals, and the correlation measure $\Lambda$ exhibit fast oscillations with frequencies $\sim \! \Delta A_{1,2}$.

Let us come back now to the apparent relative robustness of $\ket{\Psi_{01}}$ state, compared to $\ket{\Phi_{01}}$ state, to dephasing by a pair of nuclei that are strongly coupled to both of the qubits. This can be explained intuitively using the mapping of dynamics of two nuclear spin on dynamics of a pseudospin \cite{Yao_PRB06,Yao_PRL07,Liu_NJP07}. Out of four states of two nucleas spins $I\! =\! 1/2$, only $\ket{\uparrow \downarrow}$ and $\ket{\downarrow\uparrow}$ states evolve nontrivially due to dipolar interaction, and the $\mathcal{L}_{kl}$ contribution to dephasing of qubit(s) comes from dynamics in this two-dimensional subspace. One can obtain the same result by considering qubit(s) interacting with a pseudospin $\tau$ in the following way: the pseudospin is subjected to ``transverse'' field $\propto \! b\hat{\tau}_{x}$ and to a ``longitudinal field'' $\propto \! \mathcal{A}_{\pm} \hat{\tau}_{z}$ that is conditioned on the state of the qubit(s). For $\ket{\Psi_{01}}$ state, during the echo sequence the two pseudospin states conditioned on the two possible states of the qubits, experience precession along axes very close to the $z$ axis. On the other hand, for $\ket{\Phi_{01}}$ state, the evolutions of these two states involve rotations by axis close to $z$ (when conditioned on two qubits being in state $-$), and the $x$ axis (when conditioned on two qubits being in state $+$, for which longitudinal coupling of pseudospin to the qubits vanishes). Consequently, the overlap of the two states of pseudospin that determines the coherence of the qubits at echo time, is smaller in the case of $\ket{\Psi_{01}}$ state. After applying this reasononing to other cases listed in Table \ref{tab:A}, we see that a nuclear pair strongly coupled to a single qubit should create a strong ``fingerprint'' on coherence signal of $\ket{x^{+}_{01}}$ state (because $\mathcal{A}_{+} \!  \gg \! b$ and $\mathcal{A}_{-} \! =\! 0$), but {\it not} on coherence signal of $\ket{x^{+}_{1-1}}$, for which both $\mathcal{A}_{+}$ and $\mathcal{A}_{-}$ are $\gg \! b$. For two qubit coherence, pairs coupled strongly to {\it both} of the qubits should leave a strong ``fingerprint'' only on $\ket{\Phi_{01}}$ coherence, but not on all the others. A strong feature related to such a pair will however appear in $\Lambda^{\Psi}_{01}$ quantity, see Eq.~(\ref{eq:LPsi01}), because single-qubit coherences $w^{(\alpha)}_{01}$ will bear their own ``fingeprints''. 
Let us remind that when the nuclear pair is strongly coupled to {\it only one} of the qubits, the ``fingerprint'' appears for all the two-qubit coherences, but this is an effectively single-qubit contribution to $W^{\Psi/\Phi}_{ab}$, and it is going to disappear from $\Lambda^{\Psi/\Phi}_{ab}$ quantity that captures only the genuinely two-qubit aspects of dephasing.  

Based on the above discussion we expect the strong-coupling fingerprints to be more visible in $W^{\Phi}_{01}$ compared to $W^{\Phi}_{01}$, and to be least visible in $W^{\Psi/\Phi}_{1-1}$ signals, due to the fact that typical values of ``longitudinal'' fields affecting the pseudospin are larger than in $m\! = \! 0$, $1$ case.

\section{General features of electron spin echo decay due to a sparse nuclear bath} \label{sec:features}
Before we start analysing in detail the result of decoherence of entangled states of two qubits, let us recount here some known features of spin echo decay for electron spin qubits interacting with a sparse nuclear environment. We focus on these that will be relevant for discussion of two-qubit coherence results presented in Sec \ref{sec:results}.

\begin{figure}[!tb]
	\centering
	\includegraphics[width=0.9\linewidth]{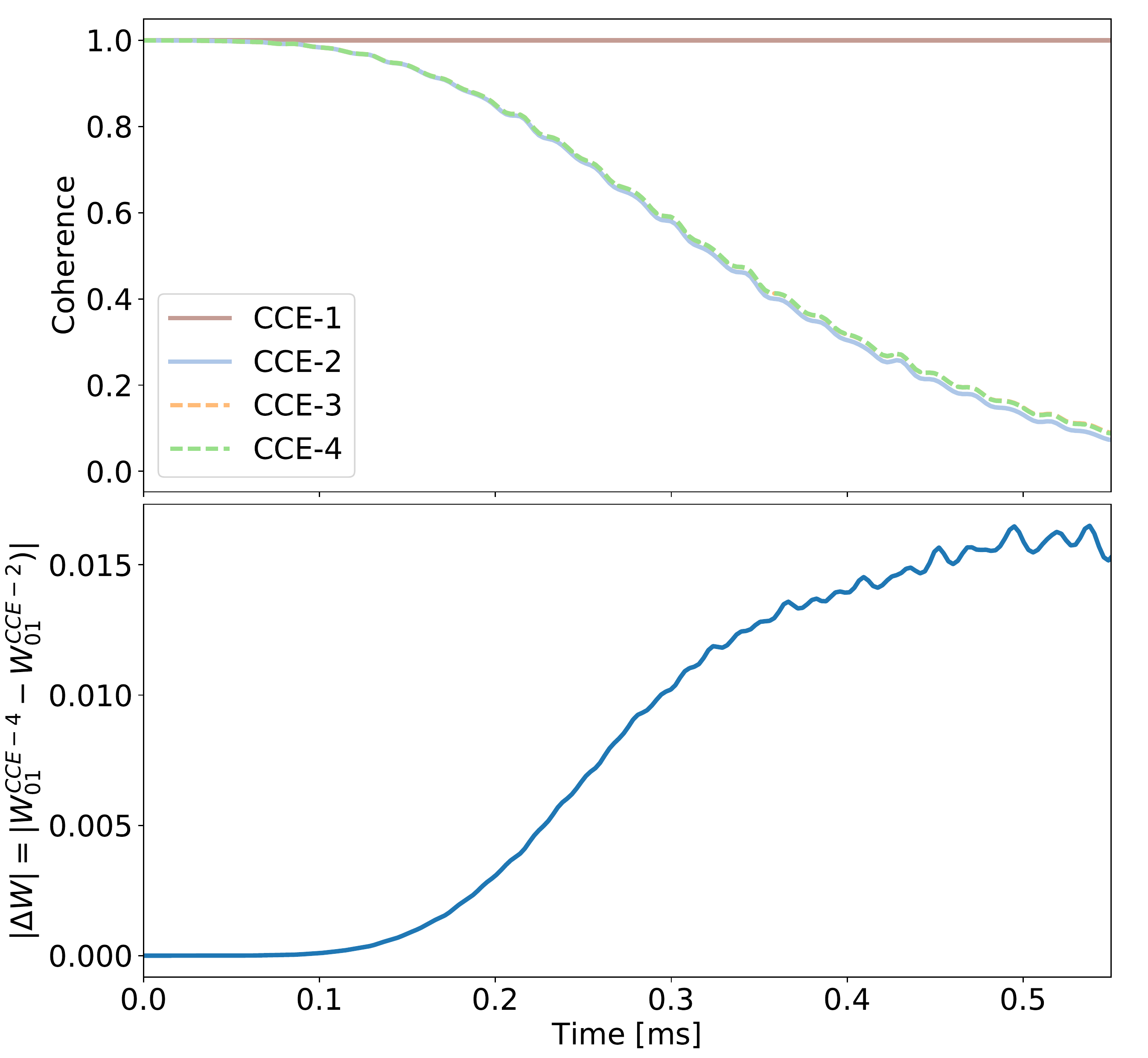}
	\caption{Convergence for CCE with $|\Delta W| \approx 0.01$. Upper panel shows coherence of a $\ket{\Psi_{01}}$ Bell state calculated with CCE-2, 3 and 4. Lower panel shows absolute value of the difference between coherence calculated up to CCE-2 and CCE-4.}
	\label{fig:errCCE-4}
\end{figure}

\begin{figure}[tb]
	\centering
	\includegraphics[width=0.9\linewidth]{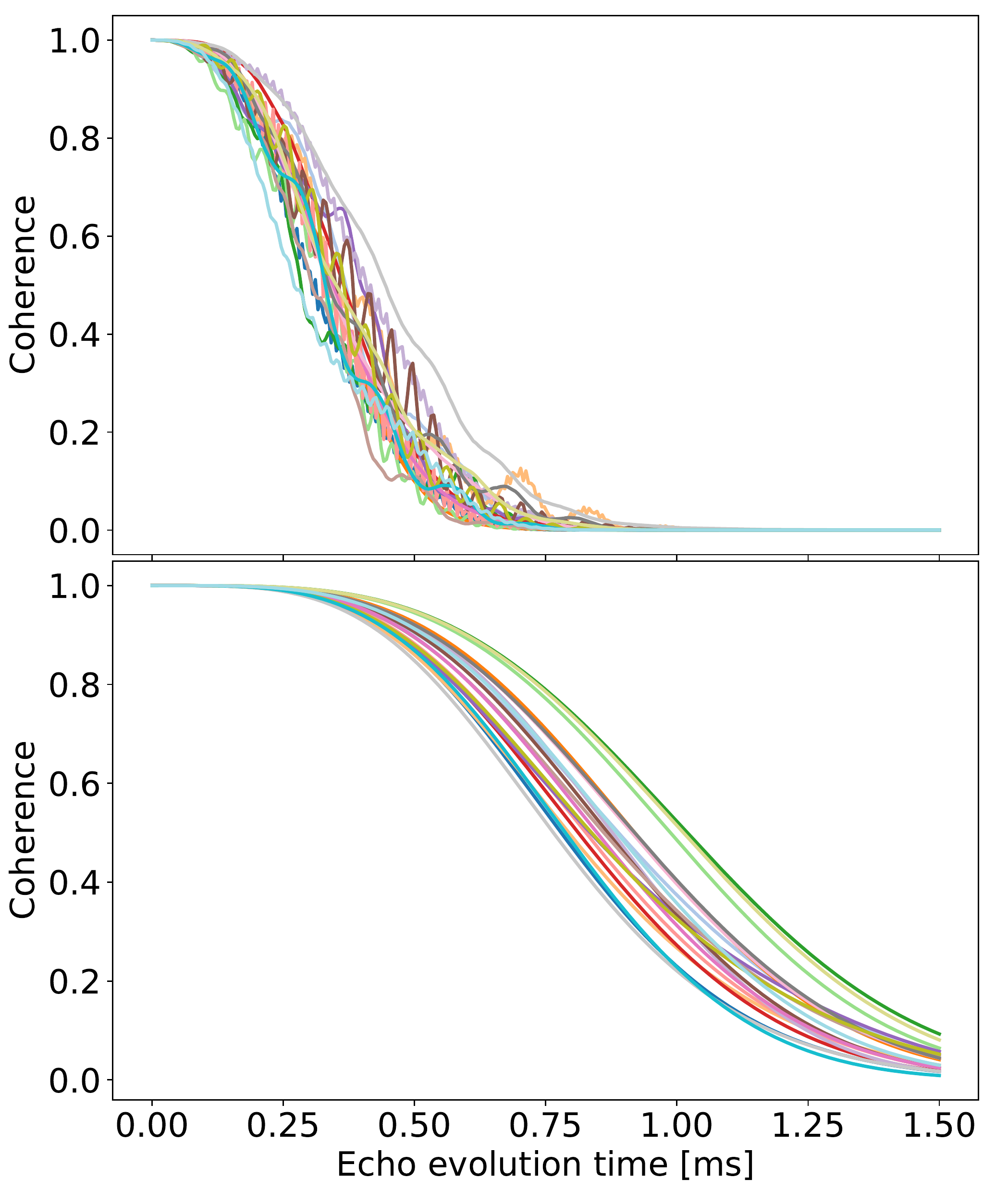}
	\caption{Figure showing single NV center decoherence due to 20 different realizations of nuclear bath at $B\! =\! 0.3$ T. The qubit is initialized in superposition of $m\! =\! 0$ and $m\! =\! 1$ states. Left panel shows a full calculation and right panel shows the same set of realizations but when strongly coupled nuclear dimers are removed within a ball of radius $r_f=4.0$ nm}
	\label{fig:realizations}
\end{figure}
 
As previously shown in \cite{Zhao_PRB12}, in case of spin echo the characteristic timescale of decay $T_{\mathrm{SE}}$, and the shape of coherence signal for times $T \! \lesssim \! T_{\mathrm{SE}}$ are well described when only dynamics of pairs of bath spins is taken into account, i.e.~the calculation is performed up to CCE-2 level. In Fig.~\ref{fig:errCCE-4} we show a calculation of two-qubit coherence performed up to CCE-4 level for magnetic field $B=0.3$ T and in presence of magnetic field gradient of 1 G/nm. The results in the presented range of times are showing convergence already at CCE-3 level (as CCE-4 and CCE-3 results are basically indistinguishable), with CCE-2 result being very similar to the converged one.
The key feature that we want to stress here is that up to the times at which coherence is an order of magnitude smaller than its initial value, the absolute error that we make by staying at CCE-2 level of calculation, given by difference between CCE-2 and CCE-4 results shown in lower panels of the Figure (in which we show typical results, as confirmed by calculations for many spatial realizations of the bath), is  $\approx 0.01$.

Another feature of decoherence caused by a {\it sparse} bath of nuclear spins is that time-dependencies of echo signals calculated for various spatial realizations of the bath can be noticeably distinct \cite{Maze_PRB08}. 
In the considered here case of $^{13}$C bath of natural concentration signals obtained for two distinct bath realizations most often differ one from another in a visible way. This is shown in the upper panel Fig.~\ref{fig:realizations}, in which we present single-spin echo calculations for 20 distinct bath realizations. While all the decays occur on similar timescale, $T_{\mathrm{SE}} \! \approx\! 0.5$ ms, they exhibit characteristic features specific to a given realization. The most visible ``fingerprints'' of certain spatial arrangement of bath spins are oscillations of the echo signal. These are caused by interaction with pairs of nuclei that are particularly strongly coupled to the qubits, discussed previously in Section  \ref{sec:strongweak}. Let us note here that when such oscillations are present, more careful analysis of the decay and also signals obtained using multi-pulse dynamical decoupling sequences, can lead to identification of specific spatial arrangements of bath spins responsible for a given ``fingerprint'' \cite{Zhao_NN11,Shi_NP14}.

In the lower panel of Fig.~\ref{fig:realizations} we show calculations done for the same 20 bath realizations, but this time we artificially remove all the spins located closer than 4 nm to the qubit. Such a ``core removal'' procedure leads not only to a visible enhancement of typical coherence time (by a factor of about $2$), but it also, unsurprisingly, leads to removal of ``fingerprints'' of strongly coupled nuclear dimers. The relative spread of coherence times also diminishes.

Finally, let us discuss briefly the influence of magnetic field gradient, used in this paper to stabilize the $\ket{\Psi_{01}}$ state against inter-qubit dipolar flip-flops, on decoherence. The gradient considered here leads to spatial dependence of nuclear Larmor frequencies, $\omega_k$, and also to appearance of $\omega^{x}_k \hat{J}^{x}_{k}$ with spatially-dependent $\omega^{x}_{k}$ that leads to tilting of precession axes of nuclei away from the $z$ axis determined by the constant component of external $B$ field. The former of these effects has straightforward influence on pair contributions to decoherence: according to Eq.~(\ref{eq:Lkl}), when $\Delta\omega \! =\! \omega_{k}-\omega_l$ is nonzero for $(k,l)$ pair, the contribution is diminished by presence of $\Delta \omega$ in the denominators (the shift of oscillations frequencies of the functions in the numerator is less important). Consequently, for large gradients we expect the contributions to decoherence of nuclear pairs oriented along the $x$ axis to be suppressed. On the other hand, the appearance of transverse splitting $\omega_x$ modifies the CCE-1 contributions (which are in any case negligible for considered here values of $B$ field), and leads to a more complicated modification of pair contributions. 
In the next Section we compare selected numerical results obtained while keeping the gradient-related corrections to CCE-1 and CCE-2 to results obtained while neglecting them. 

\section{Results for entanglement dynamics under two-qubit echo} \label{sec:results}
Most of the results presented below come from 3 different spatial realizations of the bath of uniformly distributed nuclei (enumerated as Realization 1, 2, 3). 
We will focus on the influence of the common part of the environment that makes the decoherence of $\ket{\Psi}$ and $\ket{\Phi}$ states distinct. While comparing $\Lambda^{\Psi}$ and $\Lambda^{\Phi}$ quantities defined in Eq.~(\ref{eq:Lambda}) is theoretically most natural, we will also pay close attention to the following quantity:
\begin{equation}
\delta W^{\Psi/\Phi}_{ab}=W^{\Psi/\Phi}_{ab}-w^{(1)} _{ab}w^{(2)}_{ab} \,\, .
\label{eq:deltaW}
\end{equation}
The reason is the following. The CCE-2 calculations presented below differ from the essentially exact (on the relevant timescale) CCE-4 results by at most $\approx \! 0.01$, see Fig.~\ref{fig:errCCE-4}. When $\delta W$ defined above is larger than this $\epsilon$, we can be sure that the difference between $W^{\Psi}$ and $W^{\Phi}$ calculated with CCE-2 approximation is a good approximation to the exact difference. Furthermore, since in most cases discussed below we will deal with $\delta W \! \ll \! 1$, we have
\begin{equation}
\ln(\Lambda^{\Psi/\Phi}_{ab}) = \ln\left(1+\frac{W_{ab}^{\Psi/\Phi}-{w^{(1)}_{ab}w^{(2)}_{ab}}}{w^{(1)}_{ab}w^{(2)}_{ab}}\right)\approx \frac{\delta W^{\Psi/\Phi}_{ab}}{w^{(1)}_{ab}w^{(2)}_{ab}} \,\, \
\end{equation}
as long as $\delta W^{\Psi/\Phi}_{ab}/w^{(1)}_{ab}w^{(2)}_{ab} \! \ll \! 1$, which is true as long as $w^{(q)}_{ab}$ are not very small. Since below we will focus on times of the order of half-decay time of coherence, this condition is almost always fulfilled, and $\ln \Lambda$ contains the information about the effects of the common bath that is, due to division by $w^{(1)}_{ab}w^{(2)}_{ab} $, less ``polluted'' by influences of parts of environment interacting with only one of the qubits, e.g.~fingerprints of strongly coupled nuclear pairs affecting only one of $w^{(1)}_{ab}$ discussed in Sec.~\ref{sec:strongweak}. For the case of the common part of the environment being effectively a source of Gaussian noise, according to Eq.~(\ref{eq:GaussianLambda}) we expect $\ln \Lambda^{\Psi} = - \ln \Lambda^{\Phi}$ and thus $\delta W^{\Psi} = - \delta W^{\Phi}$ (as long as both of them are small).

\begin{figure}[tbh]
	\includegraphics[width=0.8\linewidth]{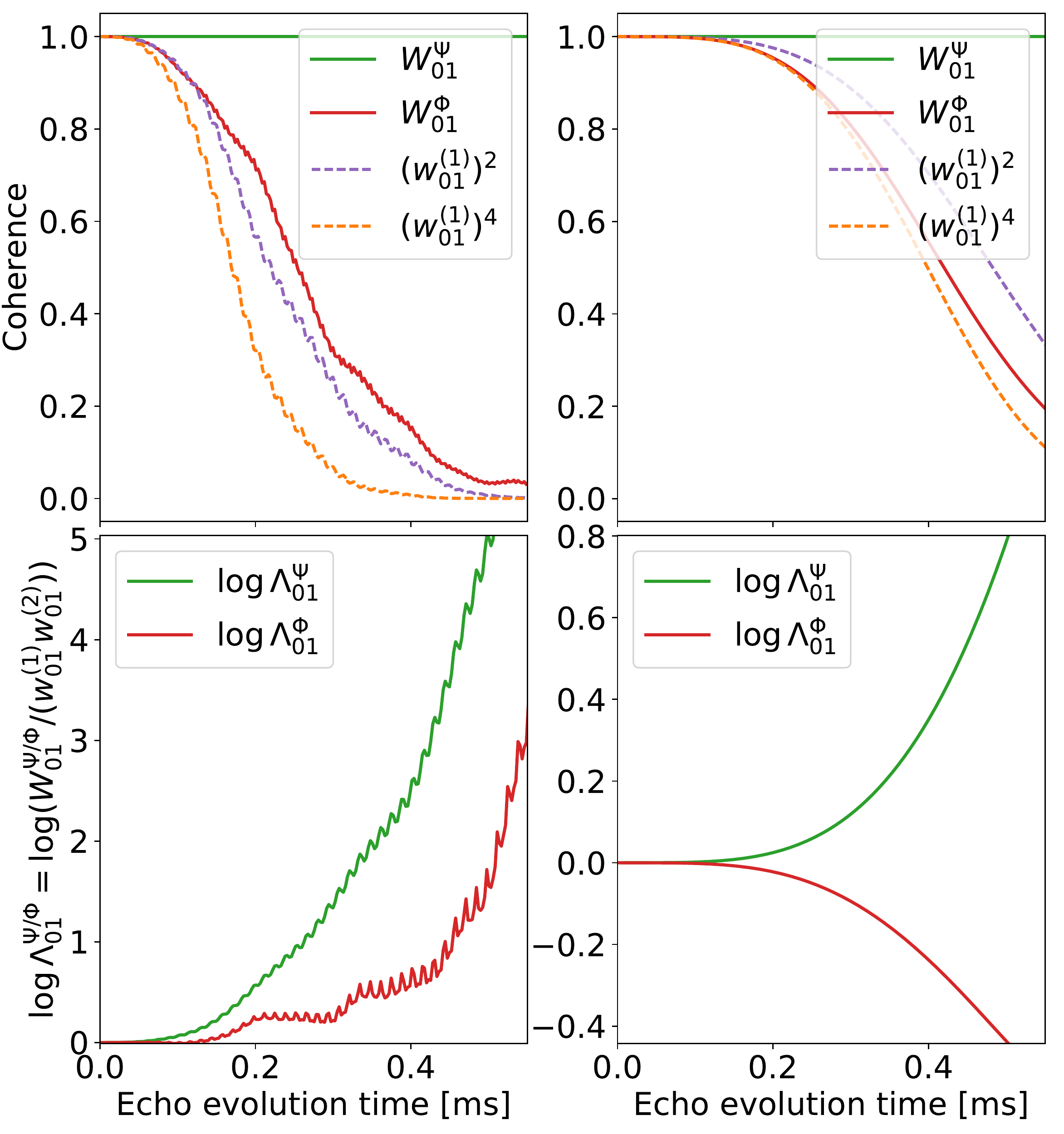}
	\caption{Decoherence of an entangled pair of NV centers located at the same spot for Realization 2, including strongly coupled nuclei (left) and also when within a ball of radius $r_f$=2 nm, those are removed (right). Upper panels shows $W^{\Psi}_{12},\,\, W^{\Phi}_{12}$ and the product of single qubit coherences: $W^1\cdot W^2$. Lower panels shows the logarithm of $\Lambda$ from Eq. \eqref{eq:Lambda}. 
	}\label{fig:NVNV0_2}
%
	\includegraphics[width=0.8\linewidth]{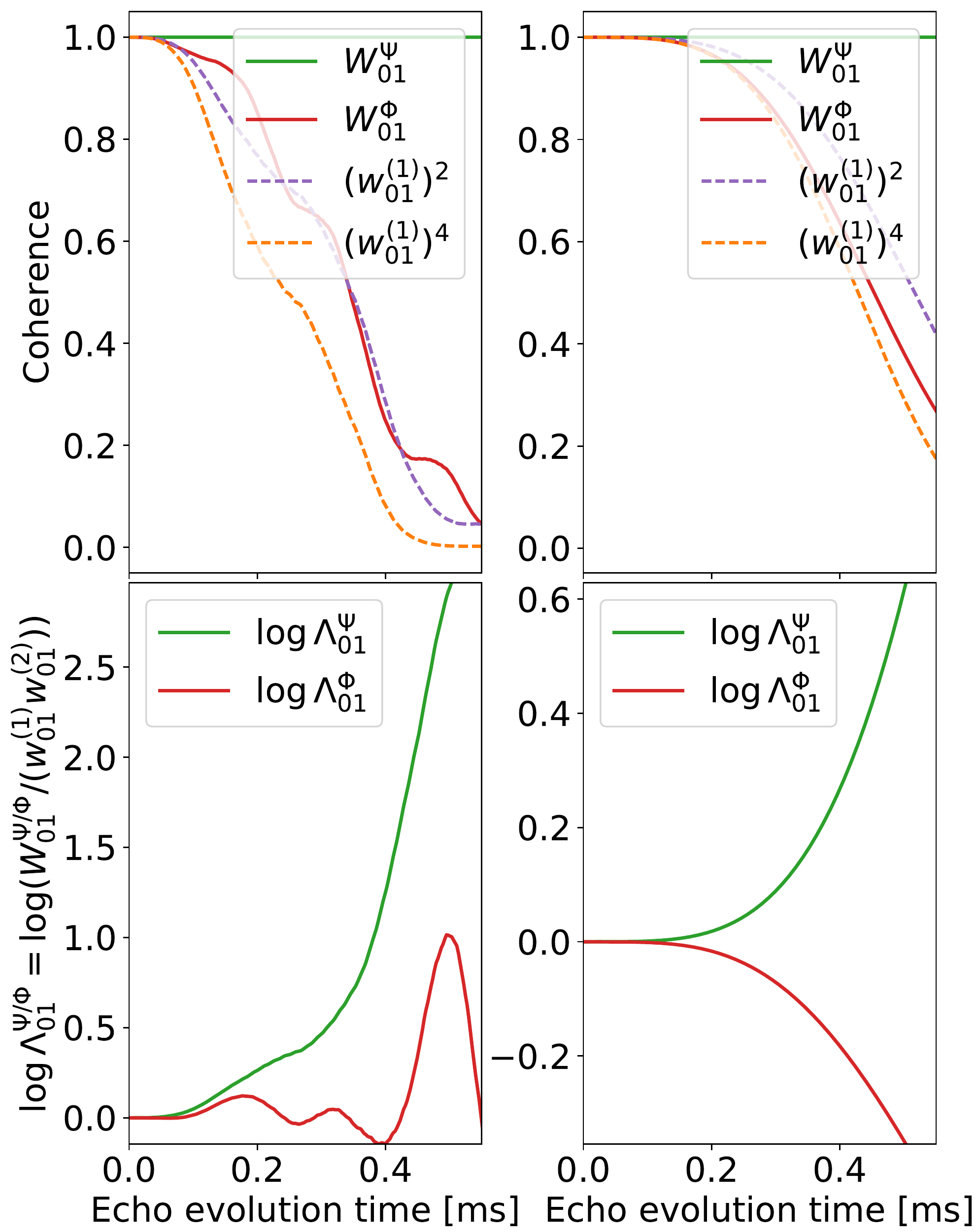}
	\caption{The same as in Fig.~\ref{fig:NVNV0_2}, but for Realization 3.
	}\label{fig:NVNV0_3}
\end{figure}

\subsection{From decoherence-free subspace for state $\ket{\Psi_{01}}$ to realistic intermediate interqubit distance}
First let us look at a somewhat artificial from experimental point of view (but theoretically interesting) case of zero interqubit distance, i.e. the case of two qubits interacting with the same common bath. As discussed in Sec. \ref{sec:correlated} the $\ket{\Psi_{ab}}$ states do not decohere then, i.e. $W^{\Psi}_{ab}(T) \! = \!1$. For $\ket{\Phi_{ab}}$ states we expect the decay that is faster than in the case of completely separate, but exactly the same environments: $W^{\Phi}_{ab}(T) \! < \! \left[w^{(1)}_{ab}(T)\right]^2$. In fact, for the Gaussian noise model of the bath, according to Eq.~\eqref{eq:GaussianLambda} we should get $W^{\Phi}_{ab}(T) \! = \! \left[w^{(1)}_{ab}(T)\right]^4$.

In Figures \ref{fig:NVNV0_2} and \ref{fig:NVNV0_3} we show the results of CCE-2 calculations for different spatial realizations of nuclear environment surrounding two qubits localized in the same place. The non-Gaussian effects are clear, as $W^{\Phi}_{ab}(T) \! \neq \! \left[w^{(1)}_{ab}(T)\right]^4$. Furthermore, we see that when strongly coupled nuclear pairs contribute to decoherence, it is not generally true that $W^{\Phi}\! < \! \left[w^{(1)}\right]^2$  -- in Fig.~\ref{fig:NVNV0_2} we see that for bath realization 2 we have $W^{\Phi}\! > \! \left[w^{(1)}\right]^2$, while in Fig.~\ref{fig:NVNV0_3} we see that for bath realization 3, $W^{\Phi}(T)$ can be smaller or larger than $\left[w^{(1)}\right]^2$ for distinct ranges of times.  However, when we remove nuclei inside a ball of radius $r_f$=2.0 nm around each qubit, we get closer to Gaussian prediction that $W^{\Phi}=(w^{(1)})^4$.
In the lower panels of these Figures we present plots of $\ln \Lambda^{\Psi/\Phi}_{01}(t)$ functions. In the full calculation, no simple relation between $\Lambda^{\Psi}$ and $\Lambda^{\Phi}$ is visible, but after the core removal the sings of $\ln \Lambda^{\Psi}$ and $\ln\Lambda^{\Phi}$ become opposite, and the results become closer to the Gaussian noise prediction of $\ln \Lambda^\Psi = - \ln\Lambda^\Phi$.

\begin{figure}[tb]
	\centering
	\includegraphics[width=\columnwidth]{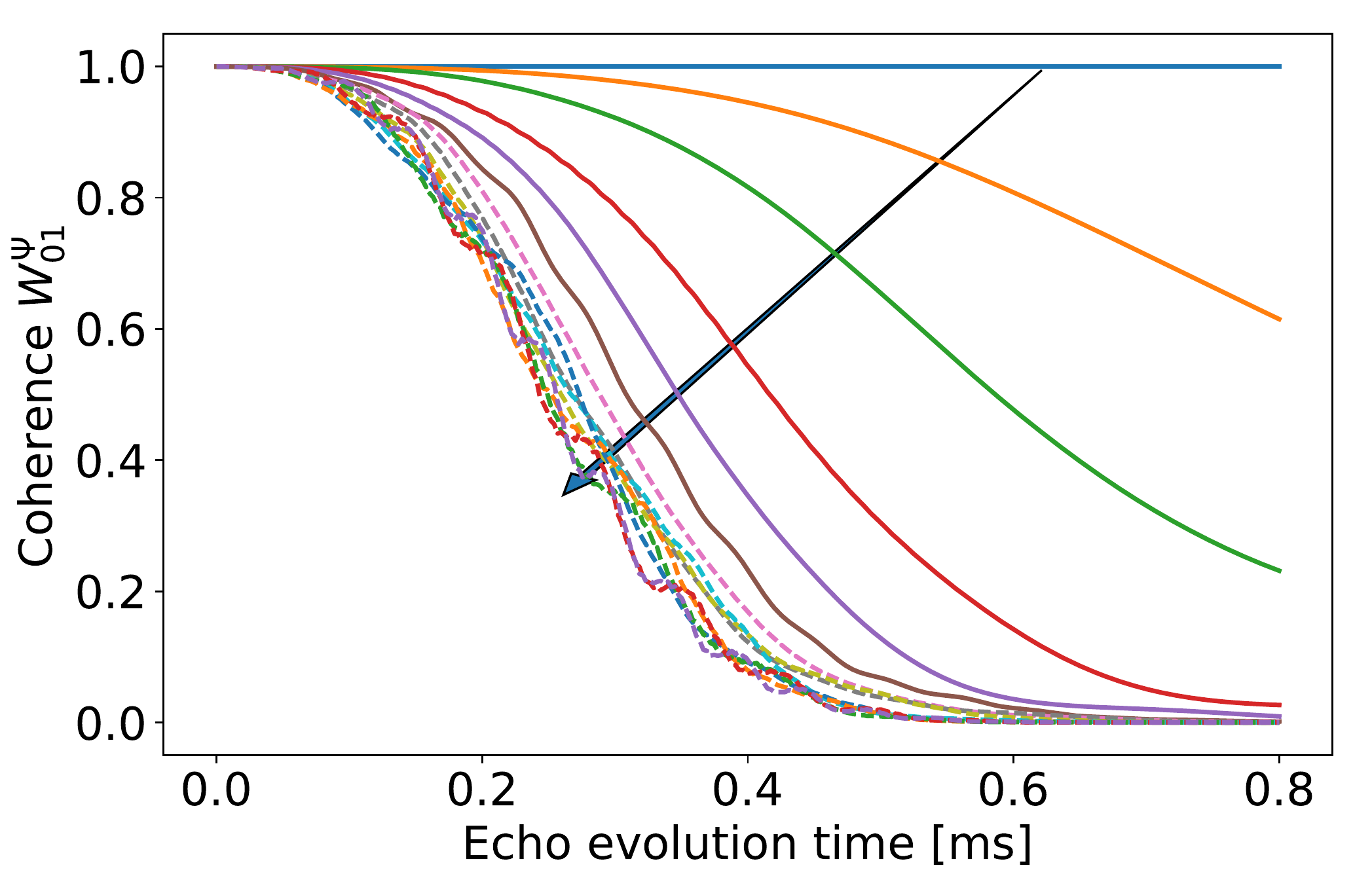}
	\caption{Decoherence of a pair of NV centers in state $\ket{\Psi_{01}}$ for Realization 1 for relatively small distances between NV centers. Solid lines correspond to interqubit distance smaller than 1 nm (specifically $d \! = \! 0$, $0.2$, $0.4$, $0.6$, $0.8$, and $1$ nm are shown) and dashed lines correspond to larger values of $d$, with maximal considered distance being $3$ nm. Arrow is pointing on curves corresponding to growing distance between NVs. 
	}\label{fig:psidfs_1}
\end{figure}

The limit of zero interqubit distance $d$ is of course unrealistic and the physically relevant question is up to what value of $d$ we can expect qualitatively similar results, i.e. $W^{\Psi}_{ab}(T)\approx 1$ on timescale $T^{\Phi_{ab}}_{E}$ for which $W^{\Phi}_{ab}(T)$ decays to less than $1/e$. As we show in Fig.\ref{fig:psidfs_1}, this common bath regime exists only for a very narrow range of $d$. For NV centers interacting with nuclear bath of natural concentration (i.e. 1.1\% of $^{13}$C isotope), the results for $d \! \approx \! 1$ nm, already look distinct from decoherence of less distant qubits and when NV centers are at larger relative distance $d> 1.0$ nm, corresponding $T_E$ times only slightly deviate from an $T_E\approx 0.5$ ms. The {\it qualitative} effect of common environment making $\ket{\Psi^{\pm}_{ab}}$ states resilient to dephasing thus disappears for inter-qubit distance $d$ of about $1$ nm for natural $^{13}$C bath. The {\it quantitative} effects of partially common environment, however, persist for larger $d$, and they are the subject of the following analysis.

\subsection{Intermediate qubit-qubit distances - non-Gaussian effects of strongly coupled nuclear pairs}
%
%
\begin{figure}[tb]
	\includegraphics[width=0.9\columnwidth]{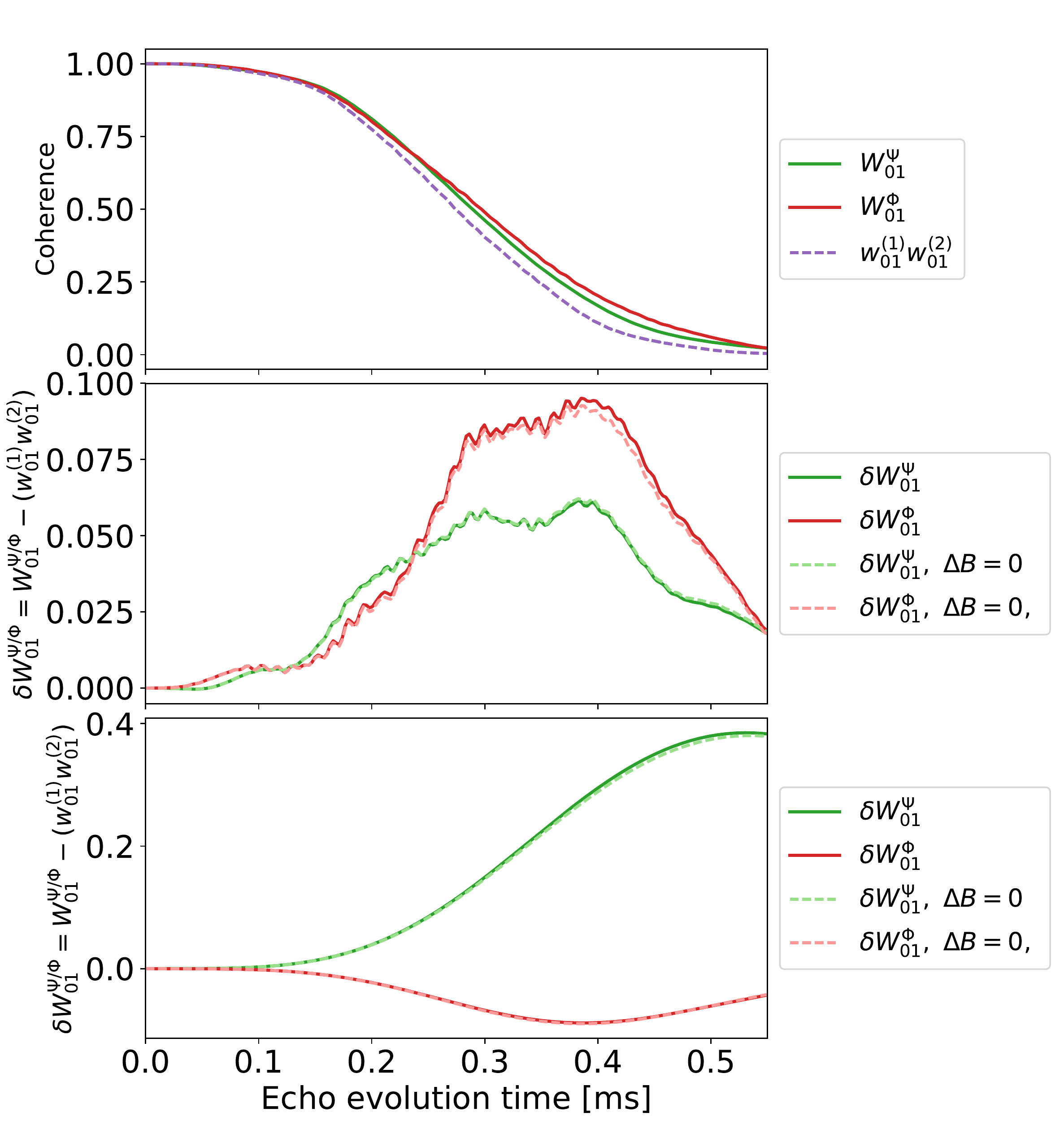}
	\caption{Decoherence of an entangled pair of NV centers separated by 1.0 nm for Realization 1. Upper panel shows $W^{\Psi}_{12},\,\, W^{\Phi}_{12}$ and the product of single qubit coherences: $w^{(1)}_{01}w^{(2)}_{01}$. Middle plot represents $\delta W$ as defined in Eq. \eqref{eq:deltaW} for both entangled states. The bottom panel shows the analogous result but with ``core'' nuclei located at $r\! < \! 2.1$ nm from each qubit artificially removed from the calculation.
	}\label{fig:NVNV1_1}
\end{figure}
\begin{figure}[bt]
	\includegraphics[width=0.9\columnwidth]{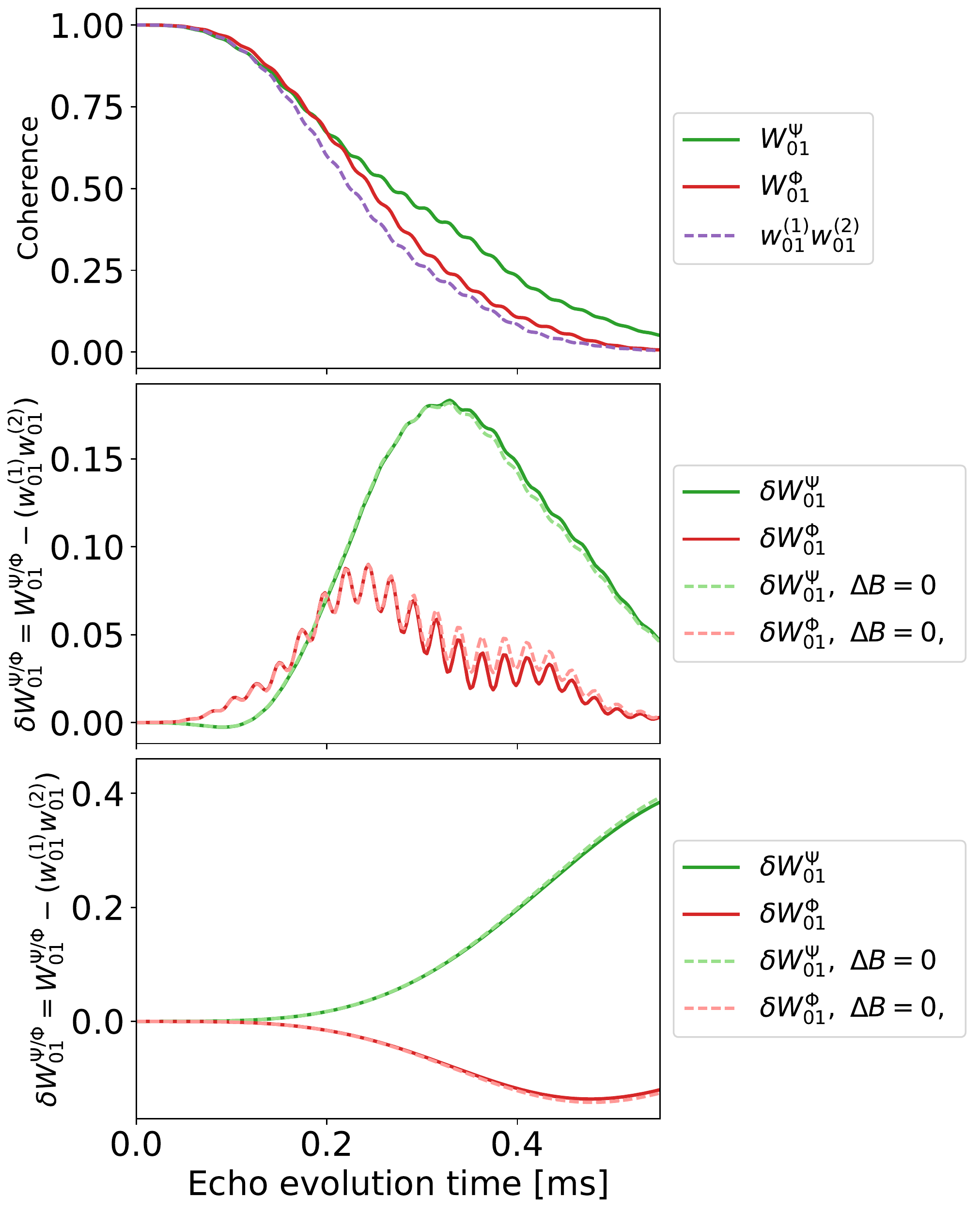}
	\caption{The same as in Fig.~\ref{fig:NVNV1_1} but for spatial Realization 2 of the bath.
	}\label{fig:NVNV1_2}
\end{figure}
In Figures \ref{fig:NVNV1_1} and \ref{fig:NVNV1_2} we show results for two spatial realizations of the bath with NV centers separated by $d \! =\! 1$ nm. First thing to note is that the contribution of common part of the bath to dephasing of $\Psi$ and $\Phi$ states is still highly non-Gaussian -- both $\delta W^{\Psi}$ and $\delta W^{\Phi}$ have the same sign. 

The second particularly interesting feature of these example results is that at such an interqubit distance the influence of the common part of the bath can appear to be similar to either positively correlated, or anticorrelated noise. According to discussion from Sec.~\ref{sec:correlated} for two qubits exposed to positively correlated (anticorrelated) Gaussian noise we should obtain $W^{\Psi} \! > \! W^{\Phi}$ ($W^{\Psi} \! < \! W^{\Phi}$). On the other hand, in Figs.~\ref{fig:NVNV1_1} and \ref{fig:NVNV1_2} we see that for the same common bath both of these relations are fulfilled at distinct timescales, with the pattern of interchanges between $W^{\Psi} \! > \! W^{\Phi}$ and $W^{\Psi} \! < \! W^{\Phi}$ being bath realization dependent -- for realization 1 the $\ket{\Phi_{01}}$ state is more coherent than $\ket{\Psi_{01}}$ for most of the times, while for realization 2 it is the $\ket{\Psi}_{01}$ state that is more resilient against decoherence for most of the considered range of times. 

In the lowest panels of Figs.~\ref{fig:NVNV1_1} and \ref{fig:NVNV1_2} we show that removal of the ``core'' of strongly coupled spins located within balls of radii $r_{f}\! =\! 2.1$ nm from each qubit makes $\delta W(T)$ curves smoother, as all the rapid oscillations caused by strongly coupled nuclear pairs are removed, and their relationship closer to the one expected for Gaussian bath, albeit only at short times. Note that, in agreement with predictions of Section \ref{sec:strongweak}, oscillatory ``fingerprints'' of strongly coupled nuclear pairs are more visible in $\delta W^{\Phi}$, see the middle panel of Fig.~\ref{fig:NVNV1_2}.
The spatial realization-dependent character of effective noise (positively correlated or anticorrelated) remains to be visible.

In these Figures we also show results obtained while neglecting the influence of magnetic field gradient on the dynamics of the bath, i.e.~using analytical formulas for CCE-2 contributions given in Section \ref{sec:twoqubitCCE}. It is clear that the gradient of $1$ G/nm, which is large enough to allow for use of pure dephasing approximation when calculating $W^{\Psi}$ at $d\! = \! 1$ nm, does not lead to modification of decoherence significant enough to affect the conclusions that we draw from all the presented results.

\begin{figure}[tb]
	\includegraphics[width=0.9\columnwidth]{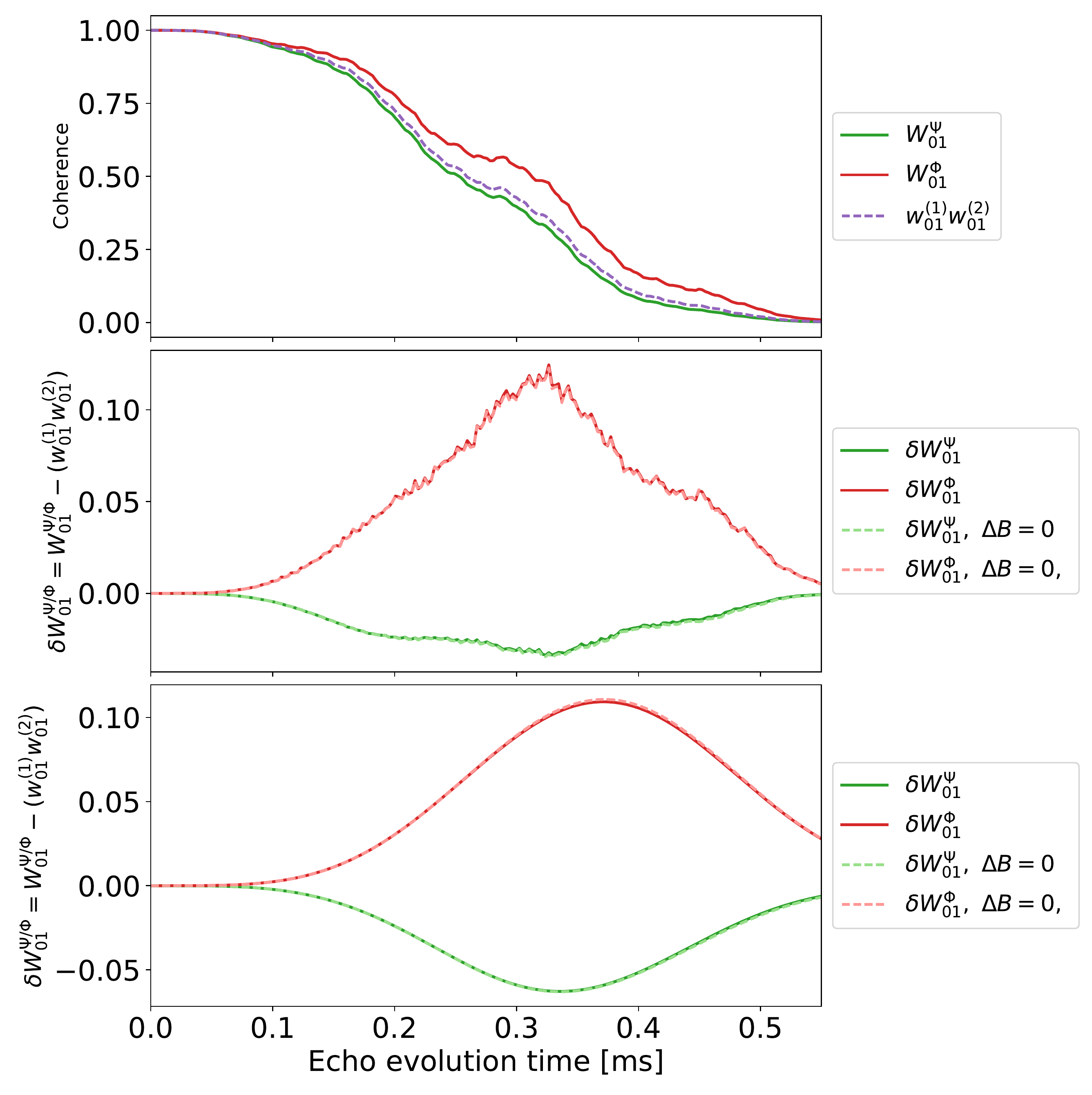}
	\caption{The same as  Fig.~\ref{fig:NVNV1_1}, but for NV centers separated by 2.0 nm and spatial Realization 1 of the bath. In the lower panel we show results with nuclei located at $r \! < \! 2$ nm from each qubit removed.  
	}\label{fig:NVNV2_1}
\end{figure}

\begin{figure}[tb]
	\centering
	\includegraphics[width=0.9\columnwidth]{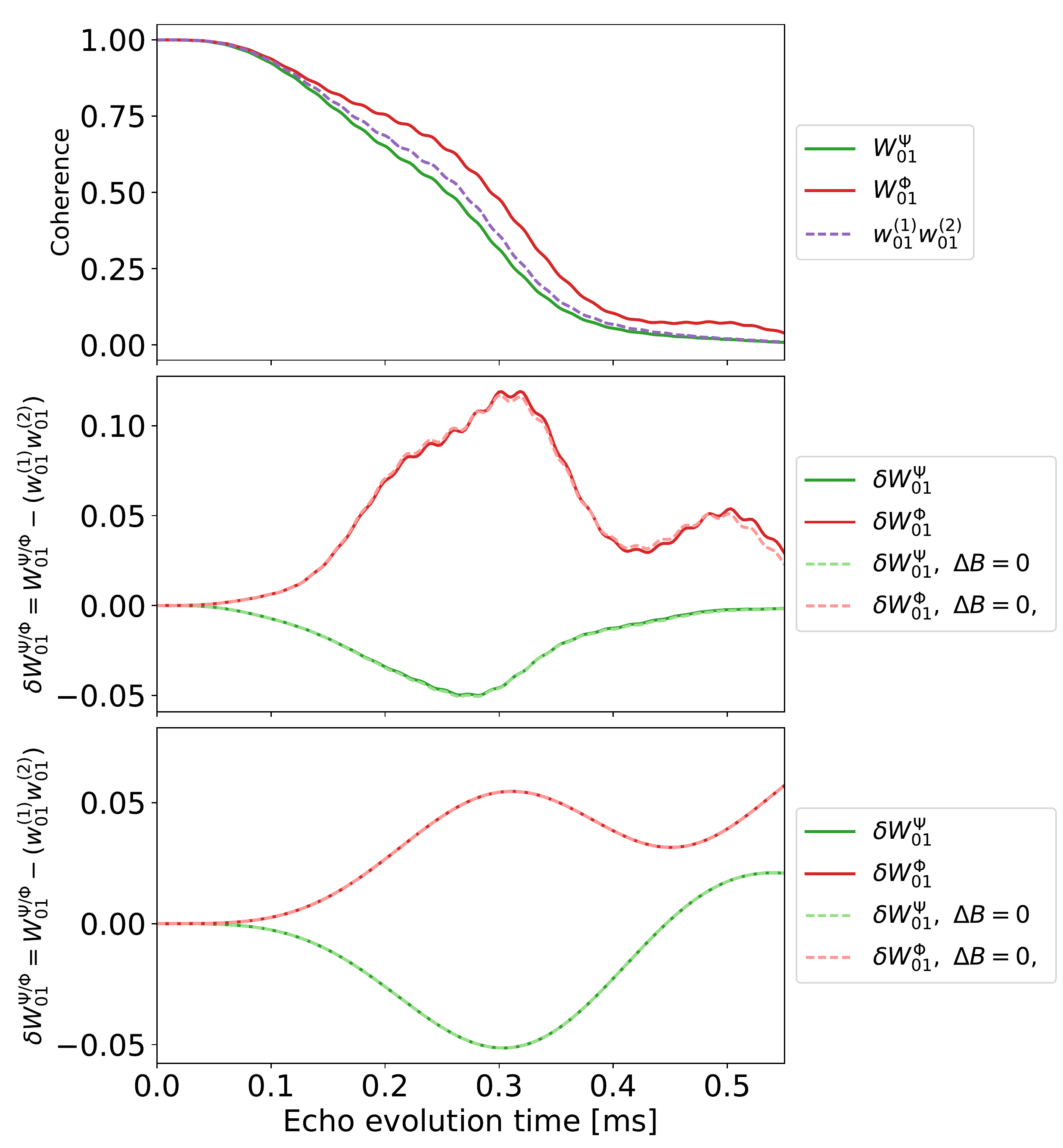}
	\caption{The same as in Fig.~\ref{fig:NVNV2_1} but for Realization 2. }
	\label{fig:NVNV2_2}
\end{figure}

In Figures \ref{fig:NVNV2_1} and \ref{fig:NVNV2_2} we show results for two spatial realizations of the bath with qubits separated by $d \! =\! 2$ nm. Coherence signals $W^{\Psi}_{01}$ and $W^{\Phi}_{01}$ are still easily distinguishable, with relative difference between them of about $30$\% on timescale of half-decay of coherence. For both shown bath realizations we have $W^{\Phi} \! > \! W^{\Psi}$ -- the transitions from effectively positively correlated to anticorrelated noise as a function of time, present in results for $d\! =\! 1$ nm, are now gone. The common part of nuclear bath appears now to act on the two qubits as common anticorrelated noise that is closer to being Gaussian -- the signs of $\delta W^{\Psi}$ and $\delta W^{\Phi}$ are opposite, and the moduli of these two quantities are of the same order of magnitude. 
As we show in the lowest panels of these Figures, removal of ``core'' nuclei within $r_{f} \! =\! 2$ nm distance from each qubit, diminishes those non-Gaussian features.

\begin{figure}[tb]
	\includegraphics[width=0.9\columnwidth]{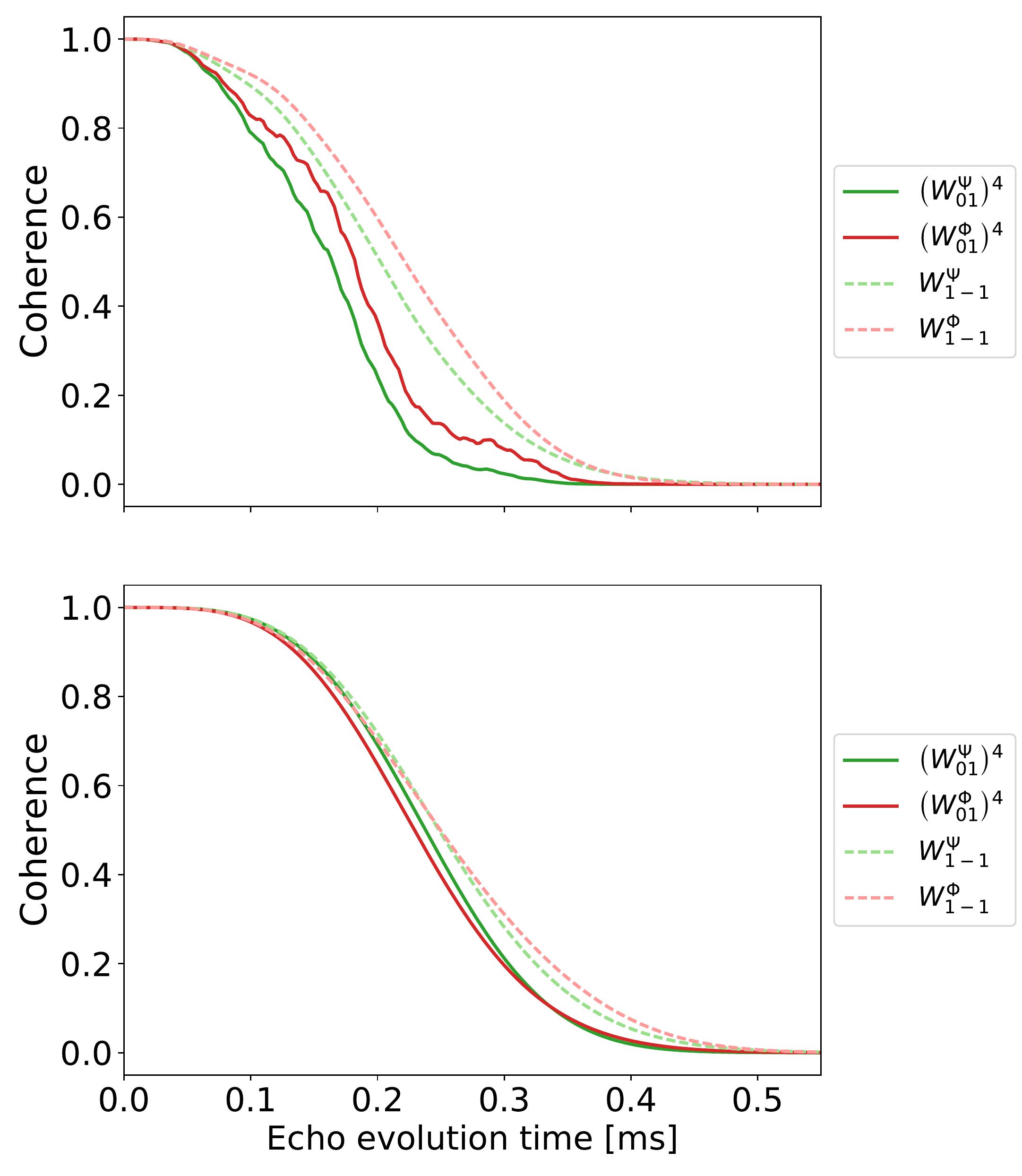}
	\caption{Decoherence of an entangled pair of NV centers separated by 2.0 nm and Realization 1: comparison between coherence between $m_s=0$ and $+1$ levels, and $m_s=+1$ and $-1$ levels. In the lower panel we show results with nuclei located at $r \! < \! 2$ nm from each qubit removed.  
	}\label{fig:double}
\end{figure}

In Fig.~\ref{fig:double} we show one example of decay of two-qubit coherence involving not $m\!= \! 0$ and $m\!= \! 1$ levels of each qubit, but $m\! =\! \pm 1$ levels. As discussed at the end of Section \ref{sec:correlated}, if the environment was a source of Gaussian noise,  we would expect $W^{\Psi/\Phi}_{1-1} \! =\! [W^{\Psi/\Phi}_{01}]^4$. As the upper panel of the Figure shows, this is clearly not the case for Realization 1 of the bath and for $d\!= \! 2$. As before, the removal of the nuclei closest to each center brings the results closer to the Gaussian case expectation, as shown in the lower panel of the Figure. It is also interesting to note, that according to discussion from Section \ref{sec:strongweak}, the fingerprints of strongly coupled nuclear pairs are absent from $W_{1-1}$ coherences - they are much smoother than $W_{01}$ results. However, as the comparison of $W_{1-1}$ and $[W_{01}]^4$ curves shows that this smoothness {\it does not} mean that the noise affecting the qubits is Gaussian. In other words, the presence of ``fingerprint'' features suggests that decoherence cannot be described with Gaussian noise model, but their absence does not prove the converse.

\subsection{Large distance - Gaussian description of correlated decoherence}
\begin{figure}[tb]
	\centering
	\includegraphics[width=0.9\columnwidth]{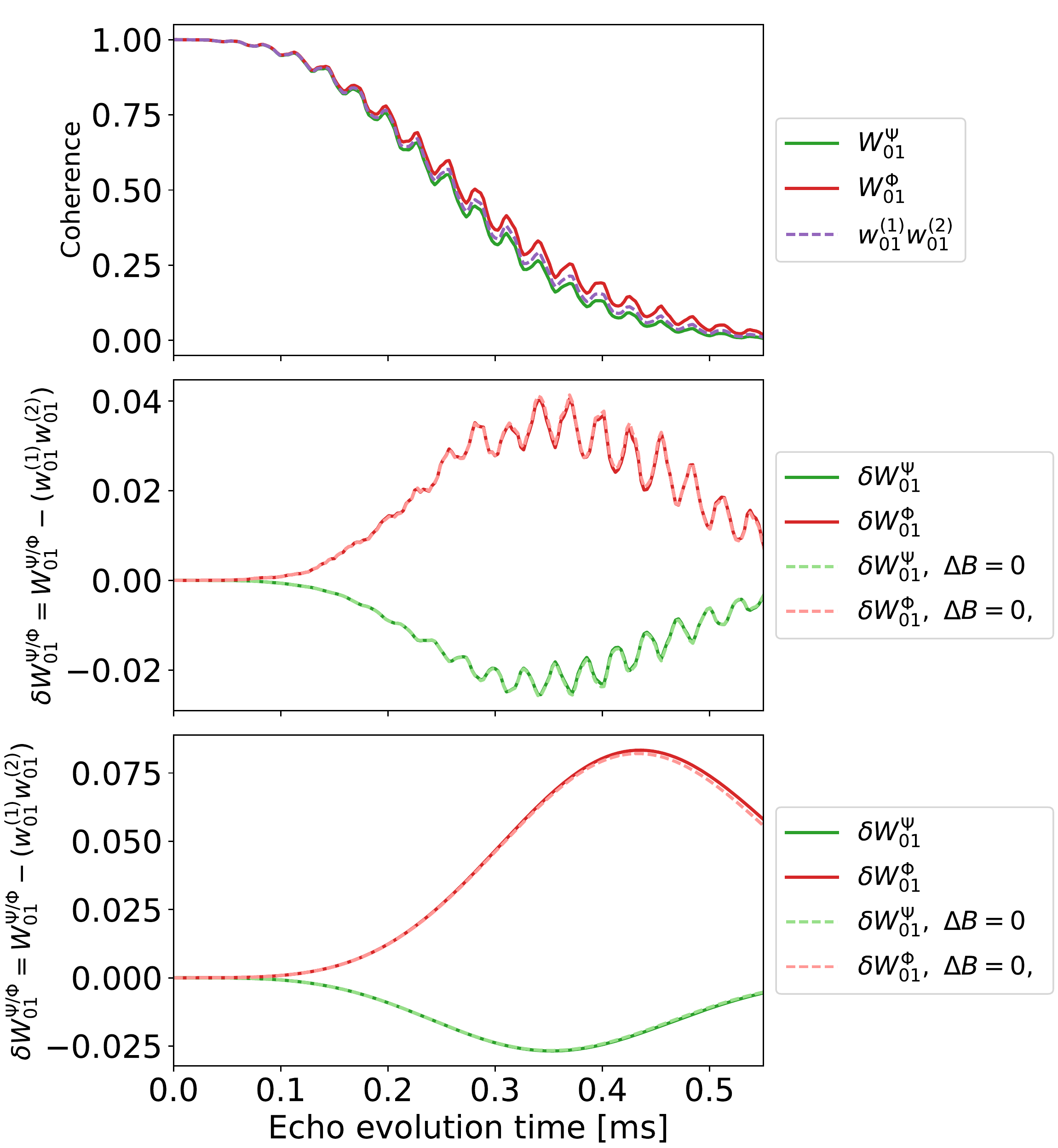}
	\caption{Same as in Fig.~\ref{fig:NVNV2_1} but for qubits separated by 3.0 nm and for spatial Realization 3 of the bath. 
	}\label{fig:NVNV3_1}
\end{figure}

When the qubits are at distances larger than 3 nm, for most spatial realizations of the environment are hardly any non-Gaussian features present in the two-qubit coherence signal. In Fig.~\ref{fig:NVNV3_1} we show examples of results obtained for spatial Realization 1. While the difference between $W^{\Psi}_{01}$ and $W^{\Phi}_{01}$ is small (at most 10\% for time comparable to half-decay characteristic time), it should be observable in experiments. The differences between the two decoherence functions are also large enough (see the middle panel of the Figure) for us to be able to claim that CCE-2 approximation correctly describes the behavior of decoherence caused by common part of the bath on timescale presented in the Figure. 

\begin{figure}[tb]
	\centering
	\includegraphics[width=\columnwidth]{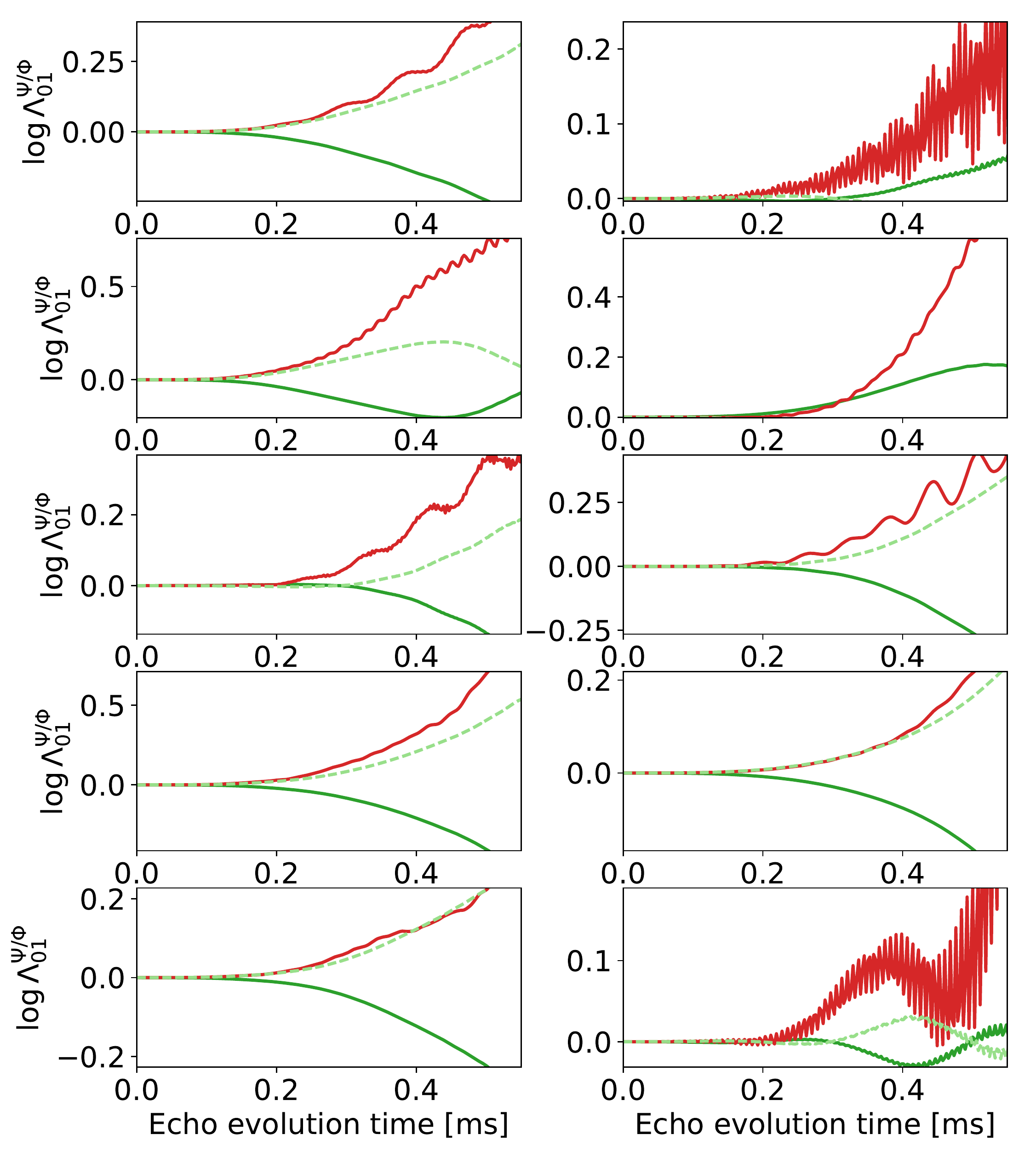}
	\caption{Correlation computed for 10 different realizations of nuclear bath with the same distance between entangled NV centers, i.e., $d$ = 3 nm. Red curve corresponds to $\log\,\Lambda^{\Phi}_{01}$, green to $\log\,\Lambda^{\Psi}_{01}$ and light green is --$\log\,\Lambda^{\Phi}_{01}$, so that we can see how far are we from classical Gaussian picture for the correlation.
	}\label{fig:ten_realiz3nm}
\end{figure}

 Figure \ref{fig:ten_realiz3nm} shows that out of ten spatial realizations of the bath, for five of them we see quantitatively Gaussian behavior of $\Lambda^{\Psi/\Phi}_{01}(t)$, for two we have a qualitatively Gaussian behavior (opposite signs of $\ln \Lambda^{\Psi}$ and $\ln \Lambda^{\Phi}$, similar magnitudes of both functions), while for three the results are strongly non-Gaussian. With interqubit distance $d$ increasing beyond $3$ nm, two things happen: the common bath effects become essentially exactly described with a Gaussian noise model, and the difference between $W^{\Psi}(t)$ and $W^{\Phi}(t)$ become very close one to another, and common bath effects become very hard to observe. 

\section{Conclusion}  \label{sec:conclusion}
We have presented a detailed theoretical description of spin-echo decoherence of an entangled pair of electron spin qubits experiencing pure dephasing due to interaction with a sparse nuclear bath, at magnetic fields high enough for decoherence to be caused by dipolar flip-flops of pairs (or larger clusters) of nuclear spins. 
We have performed calculations for NV centers in diamond, so that all the quantitative results summarized below (e.g.~the interqubit distance at which the non-Gaussian effects disappear) are specific to NV centers in diamond containing natural concentration of $^{13}$C nuclei. However, the quantitative results that we discuss here apply to other kinds of electron spin qubits coupled by dipolar interaction to a dilute nuclear environment. 

The general picture of dephasing of two-qubit coherences (that in the case of Bell states are closely related to their entanglement) is the following. The nuclear pairs that are strongly coupled to the qubits (for which $b_{kl}\ll A_k - A_l$ where $b_{kl}$ is the dipolar coupling between nuclei $k$ and $l$ and $A_{k/l}$ are qubit-nucleus couplings), cause non-Gaussian features in decoherence signal, i.e.~features that cannot be modeled by treating the environment as a source of classical Gaussian noise. These pairs have to be rather close to the qubit, as $A_{k}$ coupling decay as $\sim\!1/R_{k}^3$ with qubit-nucleus distance $R_{k}$. Pairs located farther away are weakly coupled.
For two qubits located at distance $d$ from each other, a part of the environment has significant influence on both qubits. For qubits close to each other, for interqubit distance $d$ smaller than a certain critical value $d_c$, such a common for the qubits part of the environment consists mostly of strongly coupled pairs (or, more generally, clusters). Consequently, the common the common part of environmental noise has visible non-Gaussian features. The ``witness of non-Gaussianity'' that arises naturally when considering two-qubit dephasing is the following. For Gaussian noise, the common environment contributions to decoherence of $\ket{\Psi}$ and $\ket{\Phi}$ Bell states are {\it inverse of each other}. Observation of visible deviation from such a relation is a proof of non-Gaussianity of effective noise generated by the common part of the environment -- and we have observed such deviations for two NV centers separated by $d\! < \! d_{c}\! \approx \! 3$ nm. For NV center qubits separated by $3$ nm the influence of common environment can be described in Gaussian approximation in about half of the cases, i.e.~for $50$ \% of spatial realizations of the sparse bath surrounding the two qubits. For larger $d$, the Gaussian approximation becomes exact, but decoherence signals of $\ket{\Psi}$ and $\ket{\Phi}$ states become hardly distinguishable for $d\! > \! 5$ nm. Finally, our results show that when decoherence can be modeled by assuming the environment to be a source of two partially correlated classical Gaussian noises acting on the two qubits, these noises exhibit {\it anticorrelation}, which leads to coherence of $\ket{\Phi}$ states being larger than the coherence of $\ket{\Psi}$ states.
Let us note that it should be possible to test these predictions, as pairs of NV centers separated by $\approx \! 2$ nm were observed \cite{Jakobi2016}. 

The understanding of character of common noise experienced by qubits proximal one to another is highly relevant for quantum error correction protocols, in which assume either completely uncorrelated noise affecting the qubits, thus leading to independent occurences of errors, or employ the Gaussian approximation for correlated noise \cite{Ng_PRA09}. We have presented here a quantitative analysis of decoherence caused by realistically and quantum-mechanically described common environment for a broad class of solid-state based spin qubits. We hope that the presented theory will prove useful for understanding of decoherence (and ways of counteracting it with error correction or dynamical decoupling) of multi-qubit registers based on NV centers and similar systems.

\section*{Acknowledgements}
This work is supported by funds of Polish National Science Center (NCN), grants no.~DEC-2012/07/B/ST3/03616 and no.~DEC-2015/19/B/ST3/03152. We thank Piotr Sza{\'n}kowski for his comments on the manuscript.


%

\end{document}